\documentclass{aa}  

\usepackage{graphicx}
\usepackage{txfonts}
\usepackage{booktabs}
\usepackage{latexsym}
\usepackage{array}
\begin{document} 

   \title{New composite supernova remnant toward HESS J1844-030?}

   \author{A. Petriella
          \inst{1,2,3}
          }

   \institute{Universidad de Buenos Aires, Facultad de Ciencias Exactas y Naturales, Buenos Aires, Argentina
         \and
             CONICET-Universidad de Buenos Aires, Instituto de Astronom\'ia y F\'isica del Espacio (IAFE), Buenos Aires, Argentina
             \email{apetriella@iafe.uba.ar}
         \and
             Universidad de Buenos Aires, Ciclo B\'asico Com\'un, Buenos Aires, Argentina                
            }

   \date{Received November 28, 2018; accepted April 26, 2019}

 
  \abstract
   {}
   {HESS J1844-030 is a newly confirmed TeV source in the direction of the X-ray pulsar wind nebula (PWN) candidate G29.4+0.1 and the 
complex radio source G29.37+0.1, which is likely formed by the superposition of a background radio galaxy and a
Galactic supernova remnant (SNR). Many scenarios have been proposed to explain the origin of HESS J1844-030, based 
on several sources that are capable of producing very high energy radiation.     
We investigate the possible connection between the SNR, the PWN G29.4+0.1, and HESS J1844-030 to shed light on 
the astrophysical origin of the TeV emission.}
   {We performed an imaging and spectral study of the X-ray emission from the PWN G29.4+0.1 using archival observations
obtained with the {\it Chandra} and {\it XMM-Newton} telescopes. Public radio continuum and HI data were used to derive distance constraints for the SNR that is linked to G29.37+0.1 and to investigate the interstellar medium where it is expanding. 
We applied a simple model of the evolution of a PWN inside an SNR to analyze the association between G29.4+0.1 and the radio emission 
from G29.37+0.1. We compared the spectral properties of the system with the population of TeV PWNe to investigate if HESS J1844-030 is the
very high energy counterpart of the X-ray PWN G29.4+0.1. }
   {Based on the morphology and spectral behavior in the X-ray band, we conclude that G29.4+0.1 is a PWN and that
a point source embedded on it is the powering pulsar. 
The HI data revealed that the SNR linked to G29.37+0.1 is a Galactic source at 6.5 kpc and expanding 
in a nonuniform medium. From the analysis of the pulsar motion and the pressure balance at the boundary of X-ray emission, 
we conclude that G29.4+0.1 could be a PWN that is located inside its host remnant, forming a new composite SNR.
Based on the magnetic field of the PWN obtained from the X-ray luminosity, we found that the population of electrons
producing synchrotron radiation in the keV band can also produce IC photons in the TeV band. This suggests that HESS J1844-030 
could be the very high energy counterpart of G29.4+0.1.}
   {}


   \maketitle
%

\section{Introduction}

Since the advent of ground-based $\gamma$-ray detectors such as 
the Major Atmospheric Gamma-ray Imaging Cherenkov Telescope (MAGIC),
the Very Energetic Radiation Imaging Telescope Array System (VERITAS), and the High Energy Stereoscopic System (H.E.S.S.), a
growing population of very high energy sources has being discovered in the TeV band.
Potential galactic sources of TeV radiation are supernova remnants (SNRs), pulsars and their
nebulae, high-mass binaries, and massive protostars \citep{naurois15,bosch10}. 
Even though the morphological and spectral analysis of the TeV emission can shed light on the nature of 
the source, multiwavelength observations (especially in the radio and X-ray bands) have become
a necessary and productive tool to identify the astrophysical origin of several very high energy sources. 
Nevertheless, many TeV sources still lack a clear counterpart in another spectral band and
hold the status of ``obscure''.

The H.E.S.S. Galactic Plane Survey (HGPS) has recently completed a decade of
continuum observation of the Galactic plane in the 250$^{\circ}$ to 65$^{\circ}$ longitude 
range \citep{abdalla18}. The last release of the HGPS catalog presents several TeV sources that have previously been
reported as candidates and are now confirmed as new members of the family of TeV emitting objects. 
One of the sources belonging to this class is HESS J1844-030, a newly confirmed point-like TeV source
that has previously been reported as a component of the nearby HESS J1843-033.  
The spatial coincidence between HESS J1844-030 and the radio source G29.37+0.1 indicates a probable association between them, but the complex morphology of G29.37+0.1 
has not allowed a firm identification of the astrophysical origin of the TeV radiation.  

Radio continuum observations in the direction of G29.37+0.1 first revealed an elongated
radio source with a dual jet morphology, which was classified as the candidate 
radio galaxy PMN J1844-0306 \citep{helfand89}. After the release of the Multi-Array Galactic Plane 
Imaging Survey (MAGPIS, \citealt{helfand06}), diffuse radio emission was detected around PMN J1844-0306, which was considered to 
come from a new SNR candidate. Then, the complex morphology of G29.37+0.1 in the radio band could be 
caused by the superposition of two different sources, a Galactic SNR and a radio galaxy. 

So far, the most complete study of the region was performed by \citet{caste17} (hereafter C17) using observations in different spectral bands.
Based on new radio continuum observations obtained with the Giant Metrewave Radio Telescope (GMRT), they performed a detailed morphological
and spectral study of G29.37+0.1. 
They concluded that the radio properties of the central elongated emission (referred to as an S-shaped feature) 
are similar to the emission that is observed toward other radio galaxies, while the surrounding diffuse emission (the halo) 
could be either the halo of the radio galaxy 
or a Galactic source superimposed in the line of sight. C17 judged the latter scenario to be the most probable, supported by the detection of a complex of molecular clouds at $5-6$ kpc that matches the diffuse radio emission. Then, in a Galactic scenario, 
the diffuse radio emission would be the shell of an SNR that probably interacts with the interstellar medium (ISM). 

\begin{figure}[h]
\centering
\includegraphics[width=\linewidth]{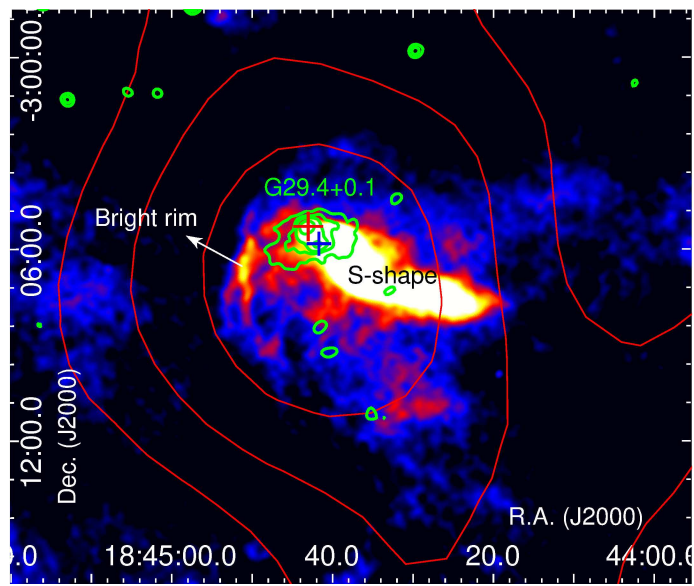}
\caption{Radio continuum emission at 20 cm (1.4 GHz) toward the radio source G29.37+0.1, extracted from the MAGPIS. 
We have marked the S-shaped feature 
(the radio galaxy) that appears surrounded by diffuse radio emission (the halo). 
The green contours are the X-ray emission from {\it XMM-Newton} in the $1.5-8.0$~keV band and show the position of 
the candidate PWN G29.4+0.1. The crosses are the X-ray point sources PS1 (red) and PS2 (blue). 
TeV emission from HESS J1844-030 is displayed with red contours. }
\label{fig1}
\end{figure}

C17 also analyzed the X-ray emission toward
the region using archival observations obtained with {\it Chandra} and {\it XMM-Newton}. 
They found diffuse X-ray emission in the direction of one of the radio lobes 
with a spectrum of nonthermal nature (see Fig. \ref{fig1}). Two point sources are located within this emission: CXO J18443.4-030520 (referred to as PS1) and CXO J18444.1-030549 (referred to as PS2). 
The excellent positional coincidence between the X-ray emission and the head of the NE lobe of the radio Galaxy point to a 
extragalactic origin, but the morphology and global spectral properties indicate that the keV emission
is likely a pulsar wind nebula (PWN) that is probably powered by PS1. This might therefore be a new composite SNR formed by the X-ray nebula (i.e., the plerion) and the diffuse
radio emission (i.e., the shell). Regarding the origin of HESS J1844-030, C17 were unable to distinguish whether the $\gamma$-rays are extragalactic and produced in the core and/or lobes of the radio galaxy or if they are the very high energy counterpart of the X-ray PWN.

This paper takes the work of C17 as a starting point to continue with the analysis of the possible connection between the different
sources that overlap HESS J1844-030. In particular, we explore the nature of the X-ray emission toward G29.37+0.1
to investigate the PWN scenario presented by C17 and the suggested association with the putative SNR traced by the radio halo. 
Following the identification adopted in the SNRcat \citep{ferrand12}\footnote{Current catalog available at \\ http://www.physics.umanitoba.ca/snr/SNRcat/}, 
we refer to the X-ray PWN candidate as G29.4+0.1. 

\section{X-ray and radio observations}

A total of six observations of the field around G29.4+0.1 are available in the database of {\it Chandra} and {\it XMM-Newton} telescopes.
C17 analyzed the two {\it Chandra} on-axis observations obtained with ACIS-I (obsID 11232 and 11801). 
In this work we also include observation 3897, which was obtained with ACIS-S toward the nearby X-ray binary AX J1845-0258. 
From {\it XMM-Newton}, C17 used observations 0602350101 and 0602350201, which targeted the nearby AX J1845-0258 in the
full-frame mode of the three EPIC cameras. A third observation is available (obsID 0046540201), but G29.4+0.1 is only fully visible
in the PN camera because the MOS1 and MOS2 cameras were operated in the small-window mode. We note that in all the three {\it XMM-Newton}
observations the source appears off-axis. For observation 0046540201, G29.4+0.1 is detected about 5$^\prime$ from the center of the PN camera.
For observations 0602350101 and 0602350201, G29.4+0.1 is detected about 9$^\prime$ and 7$^\prime$ from the 
center of the MOS and PN cameras, respectively. 

We used CIAO 4.9 and CALDB 4.7.7 to reduce and analyze {\it Chandra} data and SAS 16.1.0 and HEASOFT 6.22.1 for {\it XMM-Newton}
data. All the six observations were filtered to remove high count rate intervals. Additional filter of {\it XMM-Newton} event-files were
applied to include events with FLAG = 0 and PATTERN $\leq$ 12 and 4 for PN and MOS cameras, respectively. In Table \ref{table_X_1}
we report a summary of the observations. We show the effective exposure time of the filtered event-files 
and the background-subtracted count rate in the $1.5 - 8.0$ keV energy band extracted from the ellipse of Fig. \ref{x_imag}.

\begin{table*}[ht]
\caption{Summary of the {\it Chandra} \rm and {\it XMM-Newton} observations. 
We report the effective exposure time after filtering periods of high
count-rate and the background-subtracted count rate in the $1.5-8.0$ keV energy band from the ellipse of Fig. \ref{x_imag}, 
excluding point sources PS1 and PS2.}
\label{table_X_1}      
\centering                        
\begin{tabular}{c c c c c c}   
\hline\hline                 
Telescope & Obs-ID & Instrument & Date & Effective exposure & Count rate ($1.5-8.0$ keV) \\
          &        &            &      & time (ks) &  $\times 10^{-2}$ cts s$^{-1}$   \\
\hline                 
{\it Chandra}     & 3897       & ACIS-S         & Sept.~14, 2003 & 8.4             & 1.66  \\
                  & 11232      & ACIS-I         & Aug.~11, 2009 & 29.5             & 2.40  \\
                  & 11801      & ACIS-I         & Jun.~17, 2010 & 29.4             & 2.64  \\
{\it XMM-Newton}  & 0046540201 & PN             & March.~19, 2003 & 3.0            & 4.29 \\
                  & 0602350101 & MOS1, MOS2, PN & Apr.~14, 2010 & 34.5, 34.5, 23.3 & 1.02, 0.91, 2.85 \\
                  & 0602350201 & MOS1, MOS2, PN & Apr.~16, 2010 & 33.2, 37.5, 16.1 & 0.91, 0.80, 2.77\\
\hline                                   
\end{tabular}
\end{table*}

X-ray observations were combined with radio observations that were obtained from public surveys.
We used radio continuum data from the MAGPIS, which
maps the Galactic plane at 1.4 GHz with an angular resolution of $6^{\prime\prime}$ and typical sensitivity of 0.3 mJy. 
Spectral line observations of the HI emission at 21 cm were extracted from the VLA Galactic Plane Survey
(VGPS, \citealt{stil06}), which has angular and spectral resolutions of $1^{\prime}$ and 0.82 km s$^{-1}$, respectively, 
and an rms noise of 2 K. We also used the VGPS continuum observations at 1.4 GHz.

\section{Results}

\subsection{Morphological analysis of the X-ray emission}
\label{morf_secc}

The general morphological characteristics of the PWN G29.4+0.1 were first noted by C17. 
They presented a {\it Chandra} image that reveals an elliptical nebula from the center of which, PS1 is located offset to the NE in the direction of the minor semiaxis. In the opposite direction, they detected two protrusions 
emanating from the SW and SE. These features resemble the tongues detected toward the nebula powered by
the Geminga pulsar, where they are considered to be produced by the supersonic motion of the pulsar through the ISM \citep{caraveo03}. 

\begin{figure}[h]
\centering
\includegraphics[width=\linewidth]{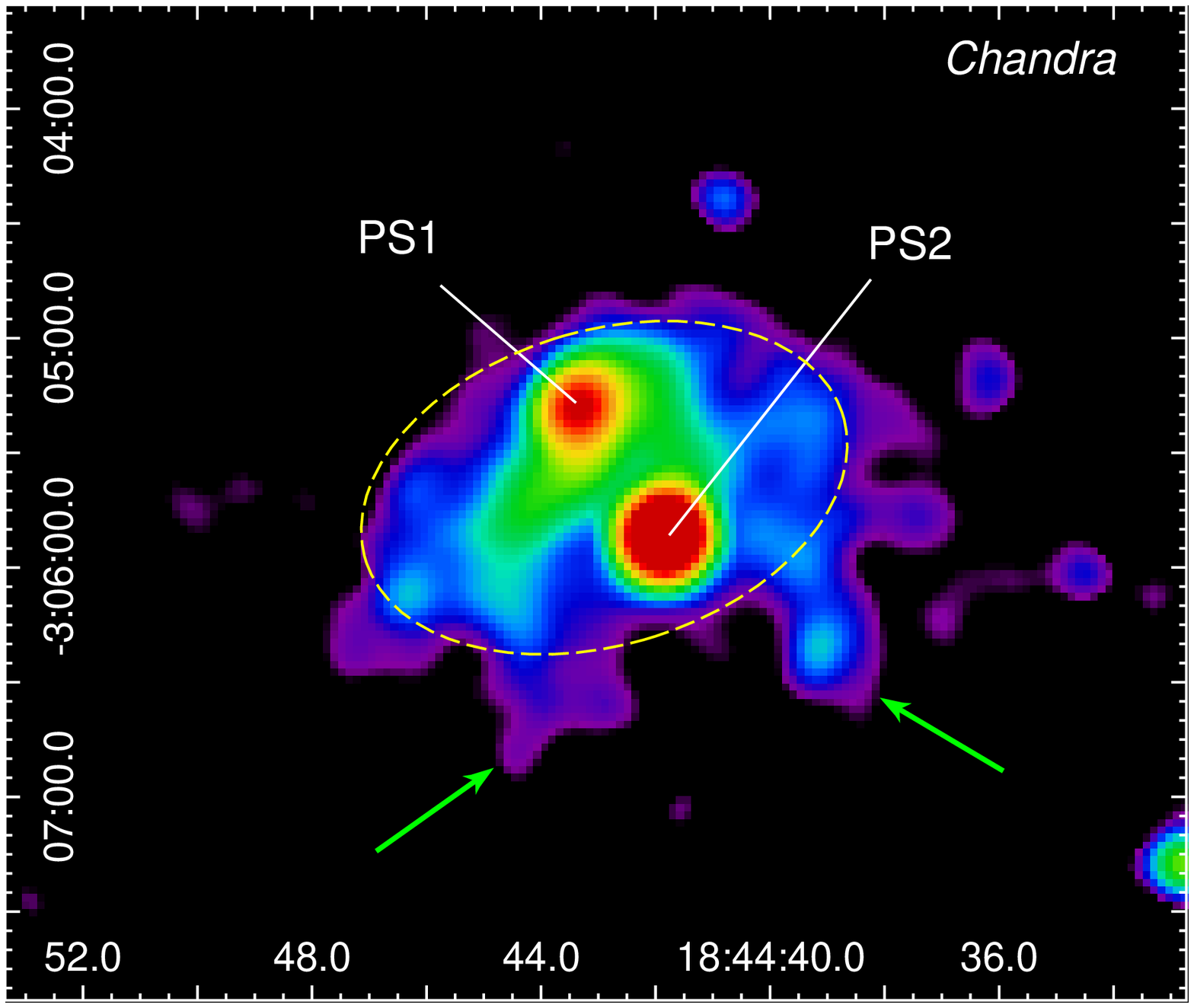}
\includegraphics[width=\linewidth]{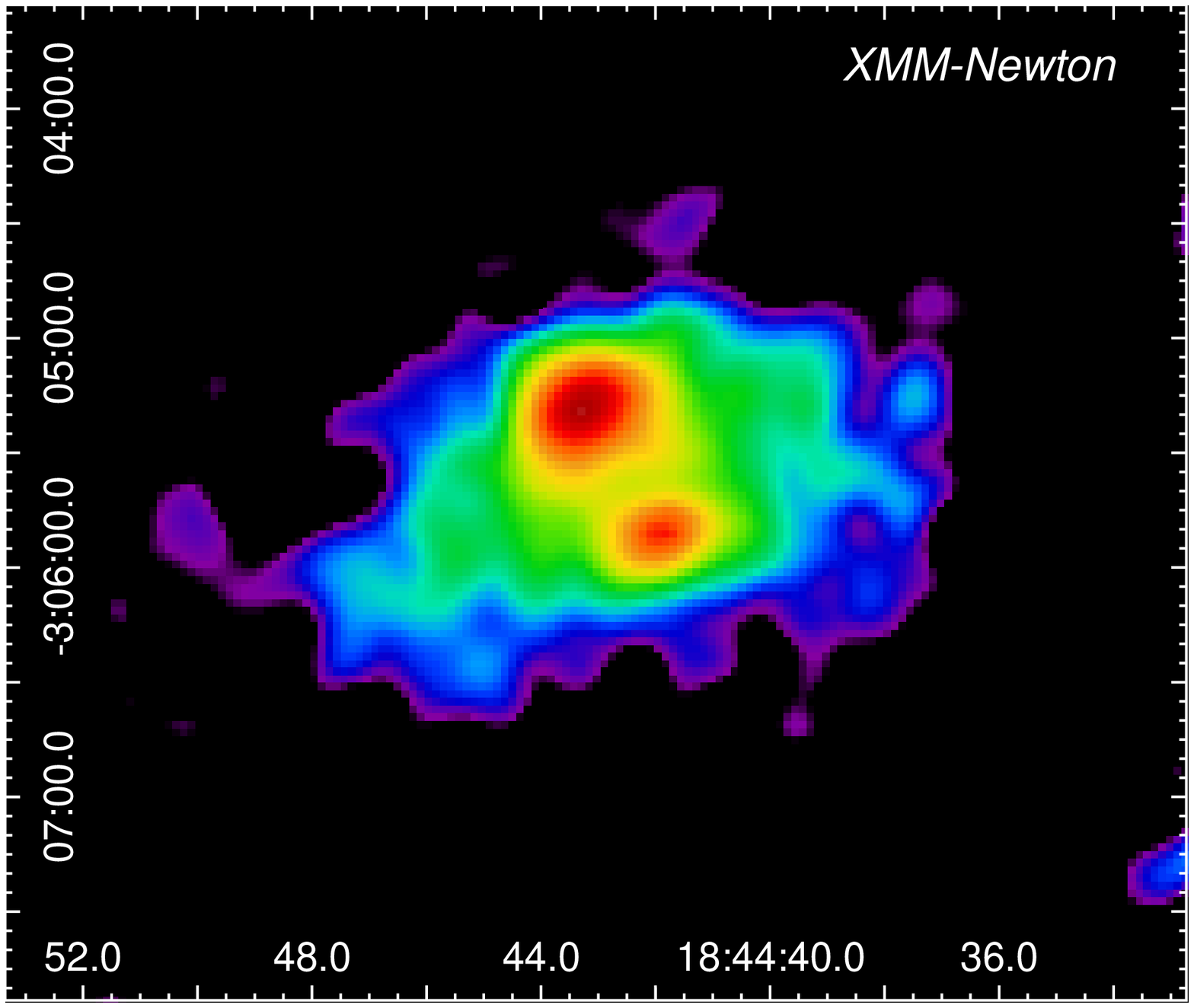}
\caption{X-ray emission in the $1.5-8.0$ keV energy band as detected by {\it Chandra} (top) and {\it XMM-Newton} (bottom).
We have marked the point sources PS1 and PS2, the ellipse used by C17 for spectral analysis, and the position of the tongues (green arrows).
The images have different color scales.}
\label{x_imag}
\end{figure}

To go further in the morphological analysis, we constructed an image of the nebula in the $1.5-8.0$ keV energy band using the long-exposure
{\it XMM-Newton} observations. No diffuse emission was detected outside this energy range, while some 
marginal emission from PS1 is seen for energies $<1.5$ keV and X-ray photons in the soft band ($<1.0$ keV) are clearly detected from PS2.  
We obtained an exposure-corrected and background-subtracted
image from the combination of six individual images corresponding to the three EPIC cameras of observations 0602350101 and 0602350201. 
The background image was constructed from the blank-sky files, and subtraction was made by scaling with the ratio
of the live exposure time of the observation and blank-sky event files.  
For comparison, we also produced an exposure-corrected mosaicked image from {\it Chandra} 
observations 11232 and 11801 using the \texttt{merge obs} task of CIAO. 
We chose the spatial binning to obtain a similar spatial scale of $2^{\prime\prime}$/pixel for the {\it Chandra} and {\it XMM-Newton} images and convolved them with a Gaussian function with $\sigma=3$ pixels.
Fig. \ref{x_imag} shows the final {\it Chandra} (top) and {\it XMM-Newton} (bottom) images. In the former, we have 
marked the position of the X-ray point sources PS1 and PS2, the ellipse used by C17 for spectral analysis and 
the tongue-like features described above (marked with green arrows).  

The X-ray morphology of the nebula revealed by the two telescopes is similar, and we can 
note two striking differences. On the one hand, PS2 appears brighter than PS1 in the {\it Chandra} image and dimmer in the
{\it XMM-Newton} image. This may be a consequence of the variability of PS2.
{\it Chandra} observation 11232 (August 2009) and {\it XMM-Newton} observations 0602350101 and 0602350201 (April 2010,
separated by only two days) likely observed PS2 in a low-energy state while {\it Chandra} 11801 (June 2010) 
detected it in a higher-energy state. 
C17 first noted the variability of PS2 and considered it a foreground source that is unrelated with the diffuse X-ray emission because its hydrogen column density is lower. 
On the other hand, we did not detect the tongues in the {\it XMM-Newton} image, which suggests 
that they may be an artifact of the binning and/or smoothing of the {\it Chandra} image.

\begin{figure}[h]
\centering
\includegraphics[width=\linewidth]{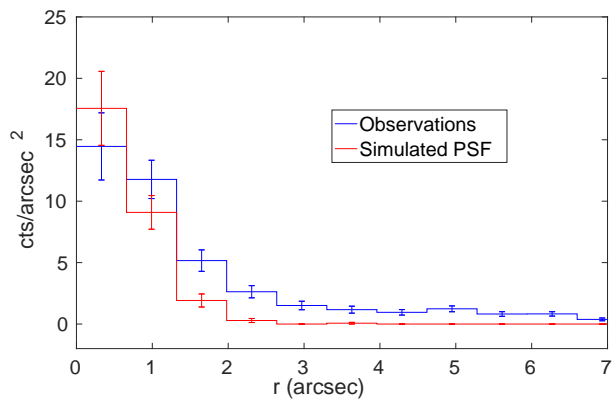}
\caption{Radial brightness profile extracted from the merged event files of {\it Chandra} observations 
11232 and 11801 (blue) and of the simulated PSF (red).}
\label{fig_radial_prof}
\end{figure}

To further explore the emission toward PS1, we took advantage of the capability of {\it Chandra}'s analysis software to
simulate the point spread function (PSF) of a point source in a particular position of the ACIS detector.  
For this purpose, we used the Chandra Ray Tracer (ChaRT) following the procedure described in 
the Ciao Science Threads. We ran one simulation each for Chandra observations 11232 and 11801, using
the spectrum of PS1 obtained by C17 as input, that is, 
an absorbed power law with $\Gamma_X = 1.75$ and $N_H = 9.6 \times 10^{22}$ cm$^{-2}$.
Then we used MARX to create an event file from the output of ChaRT, from which we constructed a full-resolution (no spatial binning was
applied) image for each observation.
From the combination of these two images, we obtained a single mosaicked image of the simulated PSF. 
We extracted a radial brightness profile in the $1.5-8.0$ keV energy band from both observed and simulated images with annular regions of width $0.7^{\prime\prime}$ centered at PS1. The results are plotted in Fig. \ref{fig_radial_prof}.
The observed brightness profile (blue line) can be explained by a point source that is embedded
in fainter diffuse emission. The bulk of X-ray photons up to $\sim 2^{\prime\prime}$ is expected to originate in the 
point source, while the diffuse emission dominates for greater distances; this is compatible with a pulsar powering a PWN.   

\subsection{Global spectral properties in the X-ray band}
\label{spec_todo}

C17 performed a spectral study of the emission from the entire nebula by fitting the spectrum they extracted from an elliptical
region in the $1.5-8.0$ keV energy band. They used an absorbed power-law model and obtained a hydrogen column 
density $N_H \sim 9.6 \times 10^{22}$~cm$^{-2}$, a photon index $\Gamma_X \sim 1.76$, and an absorption-corrected 
flux $F \sim 1.5 \times 10^{-12}$~erg s$^{-1}$ cm$^{-2}$. 
We performed a similar spectral study including {\it Chandra} observation 3897 and {\it XMM-Newton} 
observation 0046540201 (only the PN camera), which included six different observations and ten spectra that 
covered a period of seven years of observations. 
Because only few counts were extracted from these two additional observations (184 and 203 net counts in the 
$1.5-8.0$ keV band extracted from the ellipse of Fig. \ref{x_imag} for observations 3897 and 0046540201, respectively), 
we did not expect to obtain significant differences with respect to C17, but we 
have taken them into account to search for possible long-term variability of the source.
The spectra were extracted from the same ellipse as in C17, 
which is displayed in the upper panel of Fig. \ref{x_imag}, and were binned
to a minimum of 15 cts/bin. Point sources PS1 and PS2 were excluded from the extraction ellipse. 
Background spectra were obtained form nearby circular regions that were free of diffuse emission and
point sources. We used the XSPEC fitting package (version 12.9.1) and $\chi^{2}$ statistics.     

\begin{figure}[h]
\centering
\includegraphics[width=\linewidth]{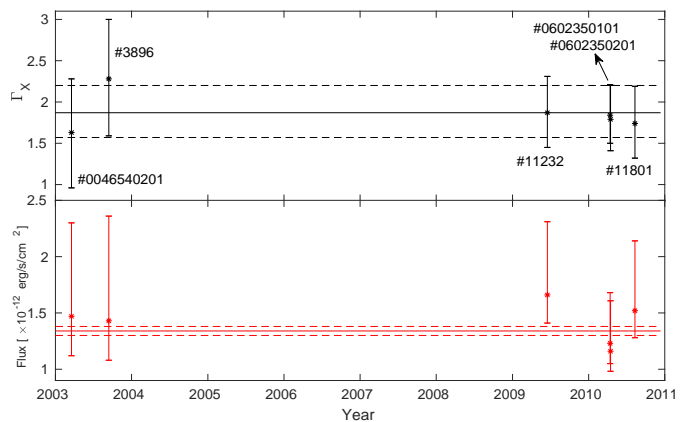}
\caption{Best-fit photon index $\Gamma_X$ (top) and absorption-corrected flux (bottom) for the simultaneous fit of 
the six X-ray observations. The horizontal filled and dotted lines represent the best-fit values and corresponding 90\% confidence
ranges that we derived from fitting the same dataset with all parameters tied together in the observations. }
\label{fig_variab}
\end{figure}

We first performed a simultaneous fit of the ten spectra for which we kept all parameters of the power-law model tied together
($N_H$, $\Gamma_X$, and the normalizations). For an absorbed power law ({\it wabs $\times$ power-law}) we 
obtained a reduced $\chi^{2}_{\nu}=1.06$ for 389 d.o.f., $N_H=(10.27_{-1.45}^{+1.65}) \times 10^{22}$~cm$^{-2}$,
$\Gamma_X=1.87_{-0.30}^{+0.33}$ and an absorption-corrected flux $F=(1.34\pm0.05)\times 10^{-12}$~erg s$^{-1}$ cm$^{-2}$.
They are in good agreement with the results of C17. We performed a second fit in which we let $\Gamma_X$ and the normalizations 
vary freely among the six observations (for {\it XMM-Newton} 0602350101 and 0602350201, the parameters of the 
three EPIC cameras within the same observation were tied). Only $N_H$ was kept frozen in all spectra. 
We obtained $\chi^{2}_{\nu}=0.96$ for 379 d.o.f. and a best-fit hydrogen column density 
$N_H=(10.15_{-1.50}^{+1.71}) \times 10^{22}$~cm$^{-2}$. The best-fit photon indexes and the derived fluxes are plotted in Fig. \ref{fig_variab}.

Fig. \ref{fig_variab} shows no variation of the photon index $\Gamma_X$ between
observations.  
The best-fit values for the flux obtained for {\it XMM-Newton} observations 0602350101 and 0602350201 
are clearly lower than for the rest of the observations, but they are still coincident when the 
confidence ranges are taken into consideration. When the same source is observed with {\it Chandra} and {\it XMM-Newton,} 
slightly different spectral results are expected, and continuous work is done in the analysis of cross-calibration between
the two telescopes\footnote{See, e.g. the {\it XMM-Newton} Cross Calibration team at https://www.cosmos.esa.int/web/xmm-newton/cross-calibration.}. In the case of G29.4+0.1, we recall that the {\it XMM-Newton} observations are considerably off-axis so that they are affected 
by the reduction of the effective area of the detector at the source position.

\subsection{Spatially resolved X-ray spectral analysis}
\label{spec_anillos}

It has been widely observed in Galactic PWNe that the photon index $\Gamma_X$ becomes softer with increasing
distance from the powering pulsar. This is a consequence of the synchrotron emission mechanism that underlies the X-ray emission because more highly energetic electrons are expected to cool faster than electrons with lower energy.
Spectral softening like this has been measured, for instance, in G0.9+0.1 \citep{porquet03}, 3C 58 \citep{bocchino01a}, G21.5-0.9 \citep{safi01},
IC 443 \citep{bocchino01b}, and KES 75 \citep{ng08}. 

\begin{figure}[h]
\centering
\includegraphics[width=\linewidth]{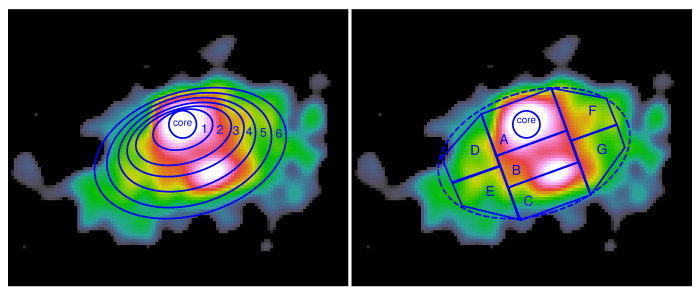}
\caption{Two sets of extraction regions we used in analyzing the spectral softening of the nebula: elliptical annular regions 
centered at PS1 (left) and polygonal regions that  are symmetrically located with respect to the minor axis 
of the nebula (right). The background image is the same as the bottom panel of Fig. \ref{x_imag}.}
\label{fig_regiones}
\end{figure} 

To further explore the PWN scenario for G29.4+0.1, we searched for variations in the photon index $\Gamma_X$ as a function of the distance
to the pulsar PS1. Because the nebula shows an elliptical morphology and PS1 appears to be off-set from its center, we defined 
annular regions with an elliptical shape that was centered at PS1. These are shown in the left panel of Fig. \ref{fig_regiones}. 
The central region (the ``core'') harbors the powering pulsar PS1, and regions 1 to 6 were defined following the orientation of
the ellipse of C17. 
Similar region definitions have been adopted to prove the spectral softening of asymmetric PWNe such as IC 443 \citep{bocchino01b} and G0.9+0.1 \citep{porquet03}. For simplicity, we excluded from the spectral analysis observations 3897 and 0046540201 because the few counts obtained in the annular regions do not contribute to the final result. 
The point source PS2 was excluded from the extraction regions.
For each region, we extracted eight individual unbinned spectra (six from {\it XMM-Newton} and two from {\it Chandra} dataset, respectively) in
the $1.5- 8.0$ keV energy band and produced a single merged spectrum for each telescope using the 
{\texttt epicspeccombine} tool of SAS and the {\texttt combine spectra} tool of CIAO.
Merging spectra can lead to different fitting results with respect to the simultaneous fit of separate spectra, which is the default procedure.
Combining observations acquired with the same telescope configuration should reduce the possibility of artifacts 
in the final merged spectrum. This is the case for {\it Chandra} observations 11232 and 11801 and for 
{\it XMM-Newton} observations 0602350101 and 0602350201. We have also performed a simultaneous fit of the individual spectra and obtained coincident results with respect to fitting the merged spectra. We report the results of the merged spectra fitting because the model parameters are slightly better constrained. 

\begin{figure}[h]
\centering
\includegraphics[width=\linewidth]{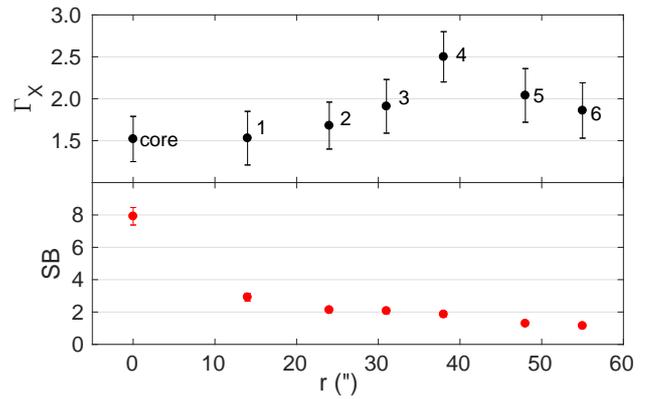}
\caption{Variation in photon index ({\it top}) and surface brightness (SB, {\it bottom}) with respect to the distance to
the putative pulsar PS1. The SB is expressed in units of $10^{-16}$ erg cm$^{-2}$ s$^{-1}$ arcsec$^{-2}$.}
\label{fig_anillos}
\end{figure}

We obtained two merged spectra for each region, which is 14 spectra for the seven annular regions. 
These were binned to a minimum of 15 cts/bin. We fit the 14 spectra simultaneously with an absorbed power law, 
freezing the absorption to the value obtained for the whole nebula in Sect. \ref{spec_todo}, 
namely $N_H=10.2 \times 10^{22}$~cm$^{-2}$. To reduce the number of model parameters, we chose a single $\Gamma_X$ and 
normalization within each region (i.e., the same for both telescopes), which should be a valid approach because these parameters vary little in the observations (see Sect. \ref{spec_todo}).
We obtained $\chi^{2}_{\nu}=1.03$ for 385 d.o.f. 
The region parameters and fitting results are shown in Table \ref{table_X_anillos}
and plotted in Fig. \ref{fig_anillos}. The parameter $r$ is the distance between the region
and PS1 and was calculated from a the EPIC-MOS {\it XMM-Newton} 
image as the mean of the distances among all the pixels of the region and the position of PS1. 

\begin{table}[h]
\caption{Best-fit parameters for an absorbed power-law model toward the seven
regions of Fig. \ref{fig_regiones} (left). The hydrogen column density was frozen to $10.2\times10^{22}$~cm$^{-2}$.
The parameter $r$ gives the mean distance between each region and PS1, as described in the text.  
The column $Counts$ shows the total number of counts ({\it XMM-Newton} + {\it Chandra}) in the $1.5-8.0$ keV energy band.  
$\Gamma_X$ is the best-fit photon index and $F$ is the absorption-corrected flux in units of $\times10^{-13}$ erg cm$^{-2}$ s$^{-1}$.  
Errors quoted are 90\%.}
\label{table_X_anillos}
\small
\centering
\setlength{\extrarowheight}{5pt}
\begin{tabular}{cccccc}
\hline\hline
 Region  &   $r$ [$^{\prime\prime}$] & Area [$^{\prime\prime 2}$]   & Counts &  $\Gamma_{X}$   &   $F$       \\
Core       &  0         & 254          &  648   & $1.52 \pm 0.27$ & $2.01 \pm 0.14 $  \\
1       &  14        & 559          &  592  & $1.53 \pm 0.32$ & $1.63 \pm 0.13$  \\
2       &  24        & 1035          &  809 & $1.68 \pm 0.28$ & $2.20 \pm 0.17$  \\
3       &  31        & 1010         &  792  & $1.91 \pm 0.32$ & $2.09 \pm 0.17$  \\
4       &  38        & 1313          &  857 & $2.50 \pm 0.30$ & $2.44 \pm 0.20$  \\
5       &  48        & 2038         &  1151 & $2.04 \pm 0.32$ & $2.62 \pm 0.20$  \\
6       &  55        & 2047         &  1089 & $1.86 \pm 0.33$ & $2.35 \pm 0.20$  \\
\hline
\end{tabular}
\end{table}

The {\it \textup{top}} panel of Fig. \ref{fig_anillos} shows that the spectrum becomes softer with increasing distance to PS1, 
at least for the inner regions of the nebula. 
The photon index $\Gamma_X$ changes from $\sim 1.5$ in the vicinity of the core to $\sim 2.5$ at about $40^{\prime\prime}$.
The {\it \textup{bottom}} plot of the figure shows that the surface brightness (i.e., the flux per unit area) 
decreases with increasing distance to PS1. Both properties are predicted by emission models of PWN in the high-energy domain and are a consequence of the underlying synchrotron radiation mechanism (see \citealt{holler12}). 

\begin{figure}[h]
\centering
\includegraphics[width=\linewidth]{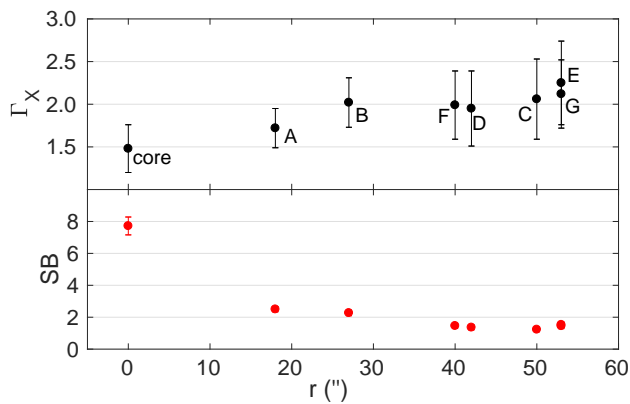}
\caption{Variation in photon index $\Gamma_X$ and surface brightness (SB, {\it bottom}) with respect to the distance to
the putative pulsar PS1 for the polygonal regions shown in the right panel of Fig. \ref{fig_regiones}. 
The SB is expressed in units of $10^{-16}$ erg cm$^{-2}$ s$^{-1}$ arcsec$^{-2}$.}
\label{fig_cajas}
\end{figure}

The lack of softening of the spectrum in the outer nebula could indicate that the particle distribution departs
from the assumed annular geometry of Fig. \ref{fig_regiones} (left). To explore this scenario, we searched for 
spectral variations using a different set of regions, which is shown in the right panel of Fig. \ref{fig_regiones}.
Regions ``core'', A, B, and C are located along the minor axis of the nebula, while regions D-E (left size of the nebula) and G-H (right
size of the nebula) can be useful to determine whether there are asymmetries in the photon index with respect to thisaxis. 
The fitting procedure (spectrum extraction, merging, and binning) was the same as before. 
We fit the 16 spectra with an absorbed power law ({\it wabs $\times$ power law}) and kept the hydrogen column density fixed to $10.2\times10^{22}$~cm$^{-2}$. 
We obtained $\chi^{2}_{\nu}=1.14$ for 334 d.o.f.. In Table \ref{table_X_cajas} we list the region parameters and 
the fitting results, which are plotted in Fig. \ref{fig_cajas}.

\begin{table}[h]
\caption{Best-fit parameters for an absorbed power-law model toward the eight polygonal
regions of Fig. \ref{fig_regiones} (right). The column definition is the same of Table \ref{table_X_anillos}.}
\label{table_X_cajas}
\small
\centering
\setlength{\extrarowheight}{5pt}
\begin{tabular}{cccccc}
\hline\hline
 Region  &   $r$ [$^{\prime\prime}$] & Area [$^{\prime\prime 2}$]   & Counts &  $\Gamma_{X}$   &   $F$       \\
Core       &  0      & 254          &  648   & $1.48 \pm 0.28$   &  $1.96 \pm 0.14$  \\
A       &  18        & 1200          &  1111  & $1.72 \pm 0.23$   & $3.01 \pm 0.19$  \\
B       &  27        & 1036          &  830 & $2.02 \pm 0.29$   & $2.35  \pm 0.18$ \\
C       &  50        & 994         &  533  & $2.06 \pm 0.47$   & $1.22 \pm 0.14$  \\
D       &  42        & 917          &  494 & $1.95 \pm 0.44$   & $1.25 \pm 0.14$  \\
E       &  53        & 882         &  516 & $2.25 \pm 0.50$   & $1.28 \pm 0.15$  \\
F       &  40        & 887         &  513 & $1.99 \pm 0.40$   & $1.30 \pm 0.14$  \\
G       &  53        & 857         &  493 & $2.12 \pm 0.40$   & $1.31 \pm 0.14$  \\
\hline
\end{tabular}
\end{table}

Fig. \ref{fig_cajas} shows a clear softening of the spectrum up to $\sim 30^{\prime\prime}$. For
$r>40^{\prime\prime}$ the spectrum becomes flatter, with a slight softening in the outermost regions E and G.
Based on the large uncertainties in $\Gamma_X$, we can only confirm a trend of spectral softening 
for the whole nebula. Finally, we do not detect asymmetries in the spectrum: the best-fit 
photon index in the left part of the nebula (regions D-E) is similar to that of the right part (regions F-G). 

For the point source PS1, C17 performed a spectral analysis using {\it Chandra} observations 11232 and 11801.
They fit a power law in the $1.5-7.5$ keV energy band and kept $N_H$ fixed to their best-fit value for the diffuse emission 
of $9.6\times 10^{22}$ cm$^{-2}$ . They obtained a photon index $\Gamma_X=1.75$. They used an extraction region
with a radius of $\sim 3^{\prime\prime}$. In Sect. \ref{morf_secc} we noted that the emission contributes
up to $\sim 2^{\prime\prime}$. Then, we restricted the extraction region to a 
radius of $2^{\prime\prime}$ and obtained a total of 100 counts in the $1.5-8.0$ keV energy band for observations 11232 and 11801.  
When the merged spectrum is fit with a power law and a hydrogen column density fixed to $10.2\times 10^{22}$ cm$^{-2}$, 
we obtain a photon index of $\Gamma_X=1.35 \pm 0.8$ ($\chi^{2}_{\nu}=0.74$ for 11 d.o.f).
The model parameters are not well constrained because of the low number counts, and the results are consistent
with those of C17. 

\subsection{Distance constraints}
\label{dist}

The X-ray emission from G29.4+0.1 alone does not allow us to establish its distance. \citet{johanson09} determined for the radio source G29.37+0.1 that the SNR candidate G29.3667+0.1000
is located between 5.2 and 15.8 kpc from the analysis of the HI absorption spectrum. 
Based on its coordinates, this source is the same as G29.37+0.1. 
C17 obtained HI absorption spectra toward the S-shaped feature, that is, the radio galaxy candidate. 
They used absorption features at negative velocities to 
suggest that this source is located beyond the solar circle and set a lower distance of 17.4 kpc, although their
analysis does not confirm or reject an extragalactic origin for this structure.
They also found a complex of three molecular clouds whose position is coincident with the halo, spanning the velocity range
between 75 and 100 km s$^{-1}$. The kinematic distance ambiguity
was resolved following the method of \citet{duval09}, which favors the near distance of $\sim 5-6$ kpc for the molecular gas complex.   
C17 did not find any spectral signature of interaction between the diffuse radio emission and
the molecular gas (such as line broadening or asymmetric profiles), but they claimed that this is not conclusive evidence to reject 
a connection between the molecular clouds and the radio halo. 

Based on the low surface brightness of the radio emission from the halo, C17 did not construct any HI 
spectra toward this structure.
We note, however, that the bright rim toward the E side of the halo (see Fig. \ref{fig1}) may be suitable for this method.
\citet{rana17} performed a systematic study of HI absorption toward SNRs and considered that a difference 
greater than 5 K between on and off positions is sufficient to obtain suitable spectra.
We defined two adjacent boxes for the on (i.e., over the continuum emission) and off (i.e., free of continuum emission) 
spectra, which are shown in Fig. \ref{figHIimag} (left) and have mean brightness temperature of 29 K and 20 K in the VGPS 1420 MHz continuum image, respectively. We note that the bright rim is not well resolved in the VGPS 1.4 GHz continuum image (shown in green contours), 
and we only detect a higher radio emission.

\begin{figure}[h]
\centering
\includegraphics[width=\linewidth]{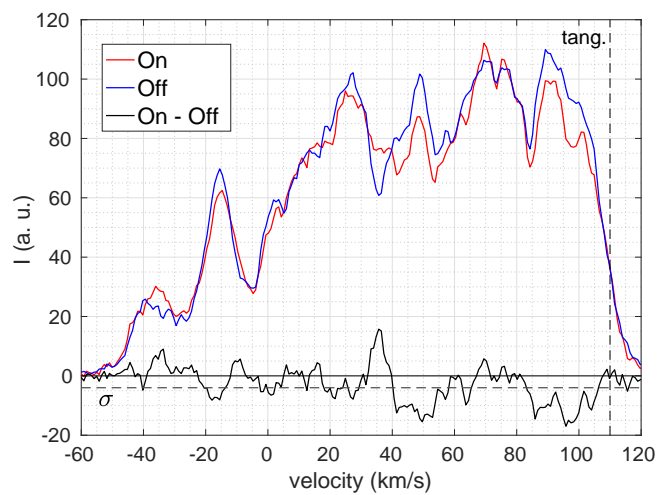}
\caption{HI on (red), off (blue) and on-off (black) spectra toward the bright rim. The vertical dashed line indicates
the velocity of the tangent point, and the horizontal dashed line is the error $\sigma$ of the on-off subtraction. }
\label{fig_HI}
\end{figure}

\begin{figure*}[ht]
\centering
\includegraphics[width=0.9\linewidth]{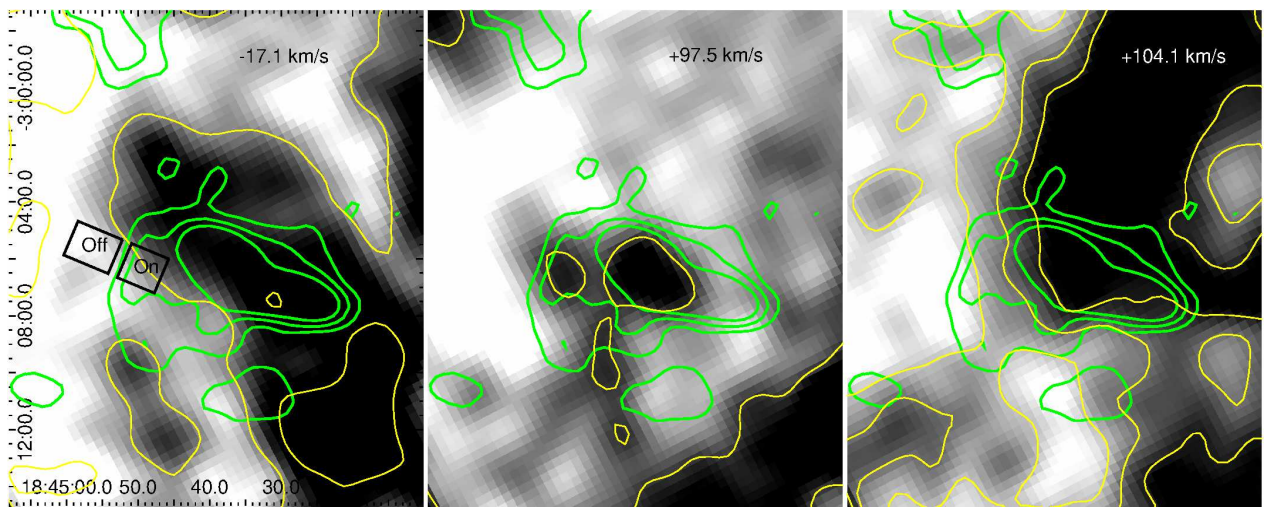}
\caption{HI channel maps (in grayscale and yellow contours) for velocities of $-17.1$, $+97.5,$ and $+104.1$ km s$^{-1}$. 
The green contours are the 1.4 GHz continuum emission extracted from the VGPS. The left panel shows the boxes used to extract
the on and off spectra. Coordinates are equatorial (J2000).}
\label{figHIimag}
\end{figure*}

We extracted the spectra from the on and off boxes and subtracted them to obtain the HI absorption spectrum, which
is shown in Fig. \ref{fig_HI}. We have marked the tangent point velocity at 110 km s$^{-1}$ and the noise level $\sigma$ of
the on-off subtraction error calculated from the VGPS $\sigma_{rms}=2$ K. 
The main features of the absorption spectrum are absorption 
at negative velocities (only the absorption at $\sim-17$ km s$^{-1}$ is clearly above the noise) 
and several absorption features at positive velocities, which seem to extend up to the tangent point. 
The last absorptions are detected at $\sim+97$ km s$^{-1}$ and $\sim+104$ km s$^{-1}$.
We inspected the HI channel map to confirm that they are not false absorptions caused
by a cloud of HI over the off position that does not extend to the on position. 
In a true absorption feature, we expect the morphology of the HI absorption to match the continuum emission of the SNR \citep{rana18}. 
In Fig. \ref{figHIimag} we show the HI channel map for $-17.1$, $+97.5$ and $+104.1$ km s$^{-1}$. 

Fig. \ref{figHIimag} shows that the morphology of the HI emission approximately matches 
the continuum emission in the direction of the bright rim only for the $+97.5$ km s$^{-1}$ channel
map. For $-17.1$ and +$104.1$ km s$^{-1}$, the absorption features are likely
caused by a HI emission in the off position, while the on position is located in a larger structure of dimmer HI emission.    
We conclude that the absorption spectrum does not present true absorption features for negative velocities and that the
last true absorption is found at $+\sim 97$ km s$^{-1}$. As the the absorption ceases before the tangent point, 
the source is likely located at the near side of the Galaxy, between the last absorption and the tangent point at $\sim 110$ km s$^{-1}$.
We used the Galactic rotation model of \citet{fich89} with a Galactocentric distance $R_0 = 8.5$ kpc and a rotation velocity
of the Sun $\Theta = 220$ km s$^{-1}$ to convert from velocity into distance and obtained a distance of $\sim 5.5-7.5$ kpc for the bright rim. 
We consider that the entire radio shell is located within this distance range. 

\subsection{Neutral hydrogen around G29.37+0.1}
\label{dens}

\begin{figure}[h]
\centering
\includegraphics[width=\linewidth]{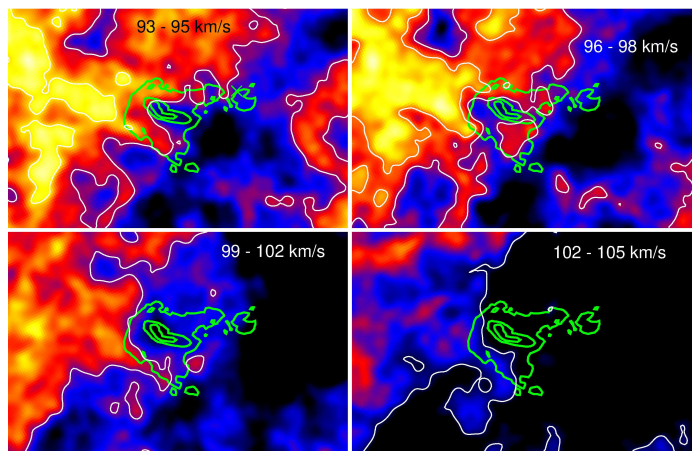}
\caption{HI channel map between 93 and 105 km s$^{-1}$. We indicate in each panel the approximate velocity interval over which 
the emission was summed. Contour levels (in arbitrary units) are 350 and 430 (upper left), 350 and 400 (upper right), 340 (bottom left), 
and 270 (bottom right). The MAGPIS radio continuum emission at 20 cm from G29.37+0.1 is shown with green contours. }
\label{HIcanales}
\end{figure}

We investigated the distribution of HI in the velocity range where G29.37+0.1 probably lies, according to the analysis
of the previous section. In Fig. \ref{HIcanales} we show a channel map of the HI emission extracted from the VGPS. 
Each panel represents the emission summed over four channels, which spans a velocity interval of $\sim 2.5$ km s$^{-1}$.
No significant HI emission is detected for velocities $>105$ km s$^{-1}$. 
The SNR is located in the interface between brighter HI emission to the left (where the radio continuum emission from the SNR
presents a more defined border) and dimmer HI emission to the right (where the radio continuum emission seems to fade). 
The intense HI emission seems to be part of a large-scale HI structure extending to the left of the SNR and 
detected along the entire velocity range shown in the figure.    
Lower HI emission is detected to the center of G29.37+0.1, forming an apparent cavity in the gas distribution.
However, we note that very low emission appears spatially coincident with the S-shaped feature, suggesting that this void could be
produced by absorption of the radio galaxy, which presents several absorption features according to C17. 
The most conspicuous evidence of interaction between the SNR and the ISM is the presence of an HI arm in the $99-105$ km s$^{-1}$ velocity
range, which protrudes from the large-scale HI emission and delineates the left and 
bottom borders of the SNR. In the Galactic rotation model, the near distance of the HI arm is $\sim 6.5$ kpc.
If the SNR has swept the neutral material, it should also be located at the same distance. Hereafter, we consider
6.5 kpc as a plausible distance for the system composed of the PWN G29.4+0.1, the putative pulsar PS1, and the radio SNR. 
    
We now discuss the density of the ambient medium into which the SNR is evolving.
The mean number density of the ISM in our Galaxy is usually considered to be $\sim 1$ cm$^{-3}$, but most of its volume
is filled with warm or hot diffuse gas with a considerably lower density.
SNRs evolving in high-density HI ($n_0 \gtrsim 1$ cm$^{-3}$) have been observed in our Galaxy \citep{park13,velazquez02}. 
Core-collapse SN explosions are expected to occur inside the stellar wind bubble that is created by the action of 
the stellar winds of the progenitor star and/or the cluster of companion stars. As a consequence, the SNR will initially 
expand in a very rarefied medium with a density that can be as slow as $n_0 \sim 0.01$ cm$^{-3}$ 
\citep{ciotti89,franco91,gaensler99,landecker99}. Because the ISM is inhomogeneous, an SNR will in a more realistic picture generally be immersed
in a medium whose density can differ considerably across the whole remnant. For instance, \citet{dubner02} studied the 
distribution of the HI around the SNR G320.1-1.2 and showed that the NW and SE borders are 
expanding in high ($\sim12$ cm$^{-3}$) and low ($\sim 0.4$ cm$^{-3}$) density gas.

Estimating the ambient density into which an SNR expands is not straightforward. In some cases, HI studies have detected SNRs
that were located inside cavities and/or superimposed to shells of neutral gas. Assuming that these structures where modeled 
by the expansion of the SN blast wave, the original ambient density is usually derived by considering that 
the swept material was distributed in the volume that the SNR occupies at the present time (see, e.g., 
\citealt{supan18,xiao12,xiao09}). In the case of G29.37+0.1, we attempt a crude estimation of $n_0$ assuming that
the SNR has swept the HI arm, which we identify as the only conspicuous structure of neutral gas interacting with the SNR. 
We integrated the HI emission between 99 and 105 km s$^{-1}$ (the velocity range where
we detect HI emission coming from the HI arm) over the area of the arm.
To calculate the HI column density, we used $N_{HI} [cm^{-2}]=1.82 \times 10^{18} \int T_B dv$, where $T_B$ is the HI brightness temperature. 
We obtained $N_{HI}\sim 9 \times 10^{20}$ cm$^{-2}$. The mass of the HI arm is $M=N_{HI}A\mu m_H$, where $A$ is the area of the arm, and 
$\mu$ is the mean atomic mass per hydrogen particle of mass $m_H$. For an ISM with 10\% of He ($\mu=1.4$), we obtain
$M \sim 1000$ M$_\odot$. If this neutral gas were originally distributed in the volume of the SNR (approximated by a sphere of 
5$^{\prime}$, which gives a radius of $\sim 8.9$ pc for a distance of 6.5 kpc), we would obtain $n_0\sim 8 $ cm$^{-3}$.
We emphasize that this result should be taken as an order-of-magnitude estimate of the ambient density into 
which the SN exploded. We have considered that the SNR has swept up only the HI arm because we cannot separate
the swept HI from the unperturbed gas in the other directions around the SNR.
Moreover, we cannot determine the effects of the stellar winds of the progenitor stars previous to the SN event that
formed the SNR. For this reason, in Sect. \ref{sedov} we explore the evolution of the SNR considering
different possible values of $n_0$ to properly account for the uncertainty in the determination of the ambient density
where the SN event occurred.  

Regarding the molecular gas, we do not find $^{13}$CO emission associated with the SNR in the Galactic Ring Survey of \citet{jackson06} 
within the velocity range of the HI arm ($99-105$ km s$^{-1}$). We do find molecular emission coincident with the radio continuum emission
at lower velocities (up to 98 km s$^{-1}$). This corresponds to a molecular cloud that was first reported by C17 
(labeled ``cloud C'' in their paper). Even if this clouds seems to delineate the right border of SNR (see Fig. 11 of C17),
we discard an interaction between them as the central velocity of the CO emission ($\sim 95$ km s$^{-1}$) is
not compatible with either the velocity of the HI arm or with the confident velocity range derived in Sect. \ref{dist}
($97 - 110$ km s$^{-1}$).

\subsection{Pulsar luminosity and spin-down energy}
\label{lumi}    

\begin{figure}[h]
\centering
\includegraphics[width=\linewidth]{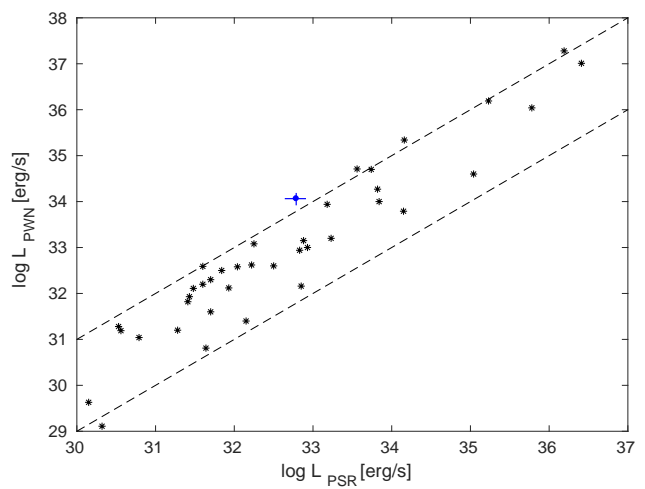}
\caption{Absorption-corrected luminosities of the extended emission (blue line; the PWN G29.4+0.1) 
and the point source PS1 (the PSR) in the $0.5 - 8.0$ keV energy band for a distance range between 5.5 and 7.5 kpc. The blue point represents the values for 6.5 kpc. The error bar corresponds to luminosities calculated between 5.5 and 7.5 kpc. The black stars correspond to a population of X-ray PWNe and their powering PSRs, and the dashed lines are upper and lower bounds defined in \citet{kargal08}.}
\label{fig_pwn_psr}
\end{figure}

After defining the distance of the system, we can calculate the luminosities of the X-ray PWN and the putative pulsar PS1 and compare them 
with a population of Galactic X-ray PWN-PSR systems. We derived the fluxes
using the results of Sect. \ref{spec_todo}. For the nebula, we considered a power law 
with $N_H = 10.2 \times 10^{22}$~cm$^{-2}$ and $\Gamma_X = 1.87$.
For the putative pulsar PS1, we used a power law with $N_H = 10.2\times 10^{22}$~cm$^{-2}$ and $\Gamma_X = 1.35$.
The calculation of the absorption-corrected flux in the $0.5-8.0$ keV energy band yields 
$F_{pwn}=2.29 \times 10^{-12}$ erg cm$^{-2}$ s$^{-1}$ and $F_{psr} = 1.21 \times 10^{-13}$ erg cm$^{-2}$ s$^{-1}$ for 
a distance of 6.5 kpc. 
In Fig. \ref{fig_pwn_psr} we plot the luminosities for 6.5 kpc, 
together with the luminosities of known PWN-PSR systems taken from \citet{kargal08}.
Error bars correspond to luminosities in the $5.5 - 7.5$ kpc distance range. 
The G29.4+0.1/PS1 system lies slightly above the fiducial upper limit defined by these authors, therefore we 
consider that the luminosities are consistent with those observed toward other galactic PWN/pulsar systems.  

Lacking the detection of pulsations from PS1, we estimate the spin-down energy $\dot{E}$ of the putative pulsar PS1 
using empirical relations reported in the literature that
relate $\dot{E}$ with the photon index and luminosity of pulsars and PWNe in different energy ranges within the keV domain.
We derived the fluxes of G29.4+0.1 and PS1 in the energy ranges using the 
best-fit model of Sects. \ref{spec_todo} and \ref{spec_anillos}. The corresponding luminosities were calculated
for a distance of 6.5 kpc. The results are reported in Table \ref{tab_Edot}.
    
\begin{table*}[h]
\caption{Estimates of the spin-down energy of the putative pulsar PS1 based on empirical relations between $\dot E$
and the X-ray emission of the pulsar and its PWN. For the photon indexes of PS1 and G29.4+0.1 we used the results
of Sects. \ref{spec_todo} and \ref{spec_anillos}, respectively: $\Gamma_{pwn}=1.87$ and $\Gamma_{psr}=1.35$. 
Luminosities in the different spectral ranges were calculated from the best-fit power-law models considering a distance of
6.5 kpc. Table references: (a) \citet{possenti02}, (b) \citet{gotthelf03}, (c) \citet{kargal08}, 
(d) \citet{seward88}, and (e) \citet{kargal13}.}
\label{tab_Edot}
\small
\centering
\setlength{\extrarowheight}{5pt}
\begin{tabular}{lclc}  
\hline\hline
Relation & Ref. &   Notes & $\dot{E}$ [erg s$^{-1}$]  \\
\hline
1) $\log{L_{psr}}=1.34\log{\dot{E}}-15.30$ &(a) & $L_{psr}$: nonthermal pulsar luminosity in $(2.0-10.0)$ keV  & $9.7 \times 10^{36}$ \\   
2) $\Gamma_{psr}=2.08-0.029 \dot{E}_{40}^{-1/2}$ &(b)& $\Gamma_{psr}$: keV photon index of the pulsar. $\dot{E}_{40}=\dot{E}/(10^{40}$ erg s$^{-1})$ & $1.6 \times 10^{37} $ \\
3) $\log{L_{psr}^{up}}=1.3\log{\dot{E}}-13.0$ &(c)&$L_{psr}^{up}$: upper limit of the pulsar luminosity in $(0.5-8.0)$ keV & $>1.7\times10^{35}$\\
\hline
4) $\log{L_{pwn}}=1.39\log{\dot{E}}-16.6$ &(d)& $L_{pwn}$: PWN luminosity in $(0.2-4.0)$ keV  & $2.5 \times 10^{36} $ \\
5) $\Gamma_{pwn}=2.36-0.021 \dot{E}_{40}^{-1/2}$ & (b)& $\Gamma_{pwn}$: keV photon index of the PWN. $\dot{E}_{40}=\dot{E}/(10^{40}$ erg s$^{-1})$  & $1.8 \times 10^{37} $ \\
6) $\log{L_{pwn}^{up}}=1.51\log{\dot{E}}-21.4$ &(e)& $L_{pwn}^{up}$: upper limit of the PWN luminosity in $(0.5-8.0)$ keV & $>5.1\times10^{36}$\\
\hline
\end{tabular}
\end{table*}

Table \ref{tab_Edot} shows that $\dot{E} \sim 1 \times 10^{37}$ erg s$^{-1}$ can be adopted as a crude estimate of
the spin-down power of PS1 that is compatible with all the empirical relations. 
We point out that according to \citet{possenti02}, their $L_{psr} - \dot{E}$ plot 
has large dispersion, and the best-fit law they obtained (relation 1) is statistically poor. They defined a more fiducial
upper limit given by $L_{psr}=10^{-18.5}\dot{E}^{1.48}$ in the $2-10$ keV band, 
which yields $\dot{E}>4.2\times 10^{34}$ erg s$^{-1}$. 
\citet{kargal13} studied a larger sample of PWN-pulsar systems (they used confirmed and 
candidate X-ray PWNe) than the other studies. 
They found some correlation between the PWN X-ray luminosity and $\dot{E}$ for low-energy pulsars, 
but a large spread for $\dot{E} \gtrsim 10^{36}$ erg s$^{-1}$. Part of this scatter is caused by distance uncertainties.
They defined the upper limit given by relation 6, which for 
G29.4+0.1 ($L_{pwn}\sim 1 \times 10^{34}$ erg s$^{-1}$ in the $0.5-8.0$ keV band) gives
$\dot{E}> 5.1 \times 10^{36}$ erg s$^{-1}$. From the inspection of their Fig. 2, 
we note that for this luminosity, their $L_{pwn} - \dot{E}$ plot is mostly populated around $\dot{E}\sim 1\times 10^{37}$ erg s$^{-1}$, 
and we take $8\times 10^{37}$ erg s$^{-1}$ as a fiducial upper limit to the pulsar spin-down energy.

Based on the previous analysis, we adopt $0.5-8.0 \times 10^{37}$ erg s$^{-1}$ as fiducial range of values of the spin-down energy of 
PS1. $1 \times 10^{37}$ erg s$^{-1}$ is a representative value that we adopt for order-of-magnitude calculations in the next sections. 

\subsection{Nebular magnetic field, synchrotron cooling time, and termination shock radius}
\label{magnetic}

To further characterize the PWN, we derived the nebular magnetic field $B$ from the synchrotron luminosity and
the spectral slope.
Assuming equipartition between particle and magnetic energy, the magnetic field that minimizes the 
total energy is obtained from Eq. 3.1 of \citet{kargal17}:
\begin{equation}
B[\mu G] = 6.7 \times 10^9 \left(\frac{\Gamma-2.0}{\Gamma-1.5}\frac{\nu_1^{1.5-\Gamma}-\nu_2^{1.5-\Gamma}}{\nu_1^{2.0-\Gamma}-\nu_2^{2.0-\Gamma}}\frac{L}{V}\right)^{2/7}, 
\label{eqB}
\end{equation}
where $L$ is the luminosity between frequencies $\nu_1$ and $\nu_2$ 
and $V$ is the source emitting volume (both expressed in c.g.s. units). We restrict our study to the $1.5-8.0$ keV 
band where synchrotron emission from G29.4+0.1 is detected.
Then, $\nu_1 = 3.6\times 10^{17}$ Hz (corresponding to $1.5$ keV) and $\nu_2 = 1.9\times 10^{18}$ Hz 
(corresponding to $8.0$ keV). We take $\Gamma=1.87$ and the flux
$F=1.34\times10^{-12}$ erg s$^{-1}$ cm$^{-2}$ (Sect. \ref{spec_todo}), which gives a luminosity  
$L = 6.8\times 10^{33}$ erg s$^{-1}$ for a distance of 6.5 kpc. When we approximate the 3D shape of the PWN with an ellipsoid that is obtained after
rotating the ellipse of Fig. \ref{x_imag} around the major axis, the volume is $V=4.5 \times 10^{56}$ cm$^{3}$.
The magnetic field of G29.4+0.1 is then $B \sim 5.6$ $\mu G$. 

The spatially resolved spectral analysis of Sect. \ref{spec_anillos} revealed
the softening of the X-ray spectrum, with $\Gamma_X$ varying from $\sim 1.5$ in the core of the nebula up to
$\sim 2.2$ in the outer regions. This spectral behavior in PWNe is attributed to the synchrotron cooling of the population
of relativistic electrons while they diffuse through the pulsar wind.
We estimated the synchrotron cooling time $\tau_{sync}$ following \citet{kargal13}: 
$\tau_{sync}\sim 38B_{\mu\rm{G}}^{-3/2} E^{-1/2}_{keV}$ kyr, 
where $E_{keV}$ is the energy of photons produced by the synchrotron mechanism 
in keV and $B_{\mu\rm{G}}$ is the nebular magnetic field in $\mu$G. 
For $E_{keV}=1.5-8$ and the magnetic field $B=5.6$ $\mu$G, we obtain $\tau_{sync}\sim 2300 - 1000$ yr.
The comparison between $\tau_{sync}$ and the pulsar age (which is equal to $t_{snr}$, the age of the host SNR) 
can shed light on the observed extent of the PWN in the X-ray band.
We found evidence that the remnant could be expanding in a high-density medium, 
in which case we obtain $t_{snr} \gtrsim 10^{4}$ yr (see Sects. \ref{dens} and \ref{sedov}). Then, 
$\tau_{sync}$ is considerably shorter than the SNR age. This may explain why G29.4+0.1 appears as a 
compact PWN around PS1 in the X-ray band, as relativistic electrons have only a short time to diffuse through the pulsar
wind before they cool through synchrotron radiation. In other words, we see emission that is produced by freshly injected electrons.  

Using the the magnetic field, we estimate the termination shock radius $R_w$ where the
pulsar wind pressure $P_{wind}$ is balanced by the internal pressure
of the nebula $P_{neb}$. For an isotropic wind, $P_{wind}=\dot{E}/(4 \pi c R_w^2)$, where $c$ is the speed of light.
Assuming equipartition, $P_{neb}=B^2/(4\pi)$ \citep{camilo06}. Taking $\dot{E}=10^{37}$ erg s$^{-1}$ and the magnetic field
derived above, $P_{wind}=P_{neb}$ yields $R_w \sim 1$ pc or $\sim 0.6^{\prime}$ for a distance of 6.5 kpc.
G29.4+0.1 presents a distorted morphology in the X-ray band, which hampers a direct comparison with the value of $R_w$ calculated
assuming spherical symmetry. We note that $R_w$ is slightly smaller than the extent of the nebula
(semiaxis $\sim 1.1^{\prime} \times 0.7^{\prime}$), and we expect it to have a distorted morphology (as the PWN):
compact ahead of PS1 (where we detect X-ray emission up to $\sim 0.3^{\prime}$) and more extended in the opposite directions.

\section{Discussion}
\label{secc_disc}

The radio source G29.37+0.1 is formed by two sources: a bright central S-shaped feature that is surrounded by a faint halo.
The former is likely a radio galaxy, while the latter could be either diffuse emission related to 
the radio galaxy or a Galactic source. Our distance analysis (Sect. \ref{dist}) shows that the halo 
is probably located at $\sim 6.5$ kpc, that is, it is a Galactic source in the foreground with respect to the radio galaxy. 
C17 claimed that if the halo were a Galactic source, it would likely be an SNR based on its shell-like appearance.
Alternatively, the halo could be the radio counterpart of the PWN G29.4+0.1, showing 
similar morphological characteristics as the SNR G327.1-1.1 \citep{ma16}.
In this source, the displacement between the radio and X-ray emissions of the PWN 
is explained by the freshly injected particles that cause the X-ray emission around the pulsar, while older material is detected 
at radio wavelength as a more extended relic PWN left behind by the motion of the pulsar \citep{temim15}. 

C17 could not determine the spectral properties of the halo because of the the low surface brightness and missing flux in the 610 MHz map. 
Nevertheless, their spectral map shows that the outer boundary of the diffuse emission 
has $\alpha \gtrsim 0.6$ ($S_\nu  \propto \nu^{-\alpha}$). 
This can be considered an indication that the nature of the halo is shell-like because the nonthermal radio continuum radiation from SNRs is steep ($\alpha \sim 0.5$)
while radio emission from plerions is flatter ($0.0 \lesssim \alpha \lesssim 0.3$) \citep{green17,gaensler06}.

C17 stated that the S-shaped feature at the center of G29.37+0.1 is likely a background radio galaxy, based 
on its jet-like morphology and distance constraints. However, \citet{abdalla18} did not completely reject the possibility that 
the S-shaped feature could be a Galactic PWN with a jet morphology and powered by a pulsar located in the core of the structure, 
similar to the PWN MSH 15-52 or the SS433\slash W50 system. In this scenario, the halo could be the host SNR
related to the S-shaped PWN and G29.4+0.1 would be a different PWN that is not associated with the radio source G29.37+0.1.  

The nature of the S-shaped feature (or part of it) as a PWN 
is supported by its position at the center of the SNR and its flat spectrum ($\alpha \sim 0.3$, C17). 
To further investigate if it could be the relic PWN linked to G29.4+0.1, we derived a crude estimate of the expected
radio flux.
The radio luminosity $L_R$ can be related with the pulsar spin-down power
as $L_R=\epsilon \dot{E}$, where $\epsilon$ is the efficiency of conversion of the pulsar power into radio emission.
From the integrated flux between $10^7$ and $10^{11}$ Hz and assuming a typical
spectral index $\alpha = 0.3$, \citet{gaensler00} derived the flux density of a PWN at 1.4 GHz  
$S_{1.4}[{\rm mJy}]=2.15 \times 10^8 \epsilon \dot{E}_{37} / D_{kpc}^2$,
where $\dot{E}_{37}=\dot{E}/(10^{37}$ erg s$^{-1})$ and $D_{kpc}$ is the distance in kpc. 
Taking $\dot{E}_{37}=1$ and a typical $\epsilon \sim 10^{-4}$ \citep{gaensler06}, we obtain $S_{1.4} \sim 0.5$ Jy for $D=6.5$ kpc.
This is on the order of the flux density at 1.4 GHz for the entire S-shaped feature ($\sim 1$ Jy, C17).
To be detectable, the radio nebula should not be too extended.  
Assuming a radius of $\sim 2.5^{\prime}$ and using the synthesized beam
of the MAGPIS ($6^{\prime\prime}.2 \times 5^{\prime\prime}.4$ at 1.4 GHz), 
we obtain a surface brightness of 0.8 mJy beam$^{-1}$, similar to the survey sensitivity 
(${\rm rms} \sim 0.7$ mJy beam$^{-1}$, \citealt{helfand06}).
Then, if the relic PWN is small (radius $<2.5^{\prime}$), it should be visible in the MAGPIS images 
and its radio emission should overlap the emission from the radio galaxy and the central part of the halo.
If the relic PWN is extended (radius $\gtrsim 2.5^{\prime}$), it remains undetectable in the 1.4 GHz MAGPIS images, and the
radio emission from the S-shaped feature can be attributed entirely to the radio galaxy. 
Further analysis is necessary to prove the presence of a PWN at the center of the SNR. In particular, the detection 
of polarized emission in the radio band could reveal the presence of the relic nebula because radio emission from PWNe is characterized by a significant fraction of polarization ($\sim 10 \%$, \citealt{gaensler06}).  

Based on its morphology, distance, and (scarce) spectral evidence, we conclude that the halo of the radio source G29.37+0.1 is 
likely a Galactic SNR, but the relic radio PWN remains unidentified. 
In the following paragraphs, we study the possible connection between this SNR, the PWN G29.4+0.1, and 
HESS J1844-030. In particular, we analyze whether G29.4+0.1 is evolving inside its host SNR and if HESS J1844-030 is the 
very high energy counterpart of the X-ray PWN powered by PS1. 
 
\subsection{Evolution of a pulsar wind nebula inside a supernova remnant}
\label{disc_pwn}

For a summary of the evolution of a PWN inside an SNR, we refer to the review paper of \citet{gaensler06}.
After a core-collapse SN event, the nucleus of the progenitor star collapses toward a compact object (neutron star or black hole) while
the stellar atmosphere is ejected, creating a blast wave that expands into the ISM.
Because the explosion is asymmetrical, the neutron star can acquire a space velocity of several hundred km s$^{-1}$.
However, the stellar ejecta initially freely expands into the ISM at a considerably higher speed
(typically, some $10^3$ km s$^{-1}$), therefore  the pulsar is expected
to be located near the center of the SNR. Because of the high sound speed inside an SNR, the pulsar wind
is confined by the gas pressure inside the remnant.
These nebulae are usually referred to as static PWNe \citep{gaensler00}.
We expect to see a symmetric PWN with a young pulsar near its center, both near the center of the remnant.

The free expansion stage of an SNR lasts for a few
thousand years and is expected to end when the mass of gas swept by the blast wave is similar to the mass of the stellar
ejecta. At this point, the SNR enters the adiabatic expansion stage (also refereed to as the Sedov-Taylor stage) which
is expected to last for some $10^4$ years. 
A reverse shock forms that decelerates and heats the expanding ejecta.
Because the ISM is inhomogeneous, the expansion of the SNR at this stage is usually asymmetric. 
As a consequence, the reverse shock is expected to return first from higher-density regions, 
reaching one side of the PWN sooner than the other, pushing the X-ray and $\gamma$-ray emissions toward regions
of lower ambient density. When it collides with the pulsar wind, the reverse shock distorts the morphology
of the nebula and forms a crushed PWN \citep{blondin01}.
At this time, the pulsar has traveled a significant distance from its birthplace, therefore we expect to see a
PWN away from the SNR center, showing a highly distorted morphology, with the pulsar possibly offset from the center of 
the nebular emission.

As the pulsar continues to move to the edge of the SNR, the sound speed of the stellar ejecta drops and
the pulsar eventually starts to move supersonically with respect to it. Now the PWN is confined by the ram-pressure of the motion 
and a bow-shock forms. The transition from a static to a bow-shock PWN is expected to occur when the pulsar has traveled
$\sim 0.68 R_{snr}$, with $R_{snr}$ the radius of the forward shock \citep{swaluw04}. 
When it is observed in radio and/or X-rays, the PWN appears within its host SNR and shows a cometary morphology
with the pulsar ahead of the emission. Representative examples of such systems are the SNRs W44 \citep{caste07,petre02} 
and IC 443 \citep{caste11,swartz15}. 
Eventually, the pulsar will cross the SNR after some tens of thousand years and continues to travel through the ISM.   
Because of the low sound speed of this medium, the motion of a young pulsar is often highly supersonic, the pulsar will 
power a bow shock, and the PWN will acquire a cometary shape with the pulsar ahead of the nebula and a trail of emission 
to the back (see \citealt{kargal17} for a gallery of bow-shock PWNe in the ISM).
The pulsar motion will take it away from the denser regions of the Galactic plane, while its spin-down energy will continue to drop. 
In the final stage of its life, an old pulsar will be found in a low-density medium 
and with insufficient energy to power a detectable synchrotron nebula.  

The PWN G29.4+0.1 is displaced from the center of the SNR and the pulsar PS1 offset with respect to the diffuse X-ray emission.
As mentioned above, this distorted morphology can be caused by a combination of pulsar motion and asymmetric
interaction of the pulsar wind with the reverse shock of the host SNR. 
If the position of PS1 ahead of the nebula indicates the direction of motion, 
it interestingly points back to the center of the SNR, supporting a common origin for both of them.
Then, the SNR linked to the radio source G29.37+0.1 and the X-ray PWN G29.4+0.1 would form a new composite SNR.  
This scenario is analyzed in the following paragraphs using the properties of 
the PWN, the pulsar, and the surrounding medium we obtained in previous sections, together with a simple model
that describes the structure and evolution of a PWN inside an SNR during the adiabatic stage.  

\subsection{Sedov solution for the SNR}
\label{sedov}

The Sedov solution for the expansion of an SNR during the adiabatic stage is derived by assuming an 
SN explosion that instantaneously releases an amount of energy $E_0$ into a homogeneous medium of density $\rho_0$.
About 99\% of the total explosion energy of an SN
is carried away by the neutrinos, and only the remaining fraction is injected into the ISM as mechanical energy, which is assumed
to have a canonical value $E_0\sim 10^{51}$ erg.
The evolution of the forward shock at $R_{snr}$ is given by \citep{vink12}
\begin{equation}
R_{snr} = \xi_0 \left(\frac{E_0}{\rho_0} \right)^{1/5} t^{2/5},
\label{eq1}
\end{equation}
where $\xi_0=1.15$ and $\rho_0 = \mu m_H n_0$, being $\mu$ the mean atomic mass per H particle, $m_H$ the mass
of the H atom, and $n_0$ the number density of the medium where the SNR expands. Hereafter, we assume an ISM
with 10\% of He, which leads to $\mu = 1.4$. 

\begin{figure}[h]
\centering
\includegraphics[width=0.8\linewidth]{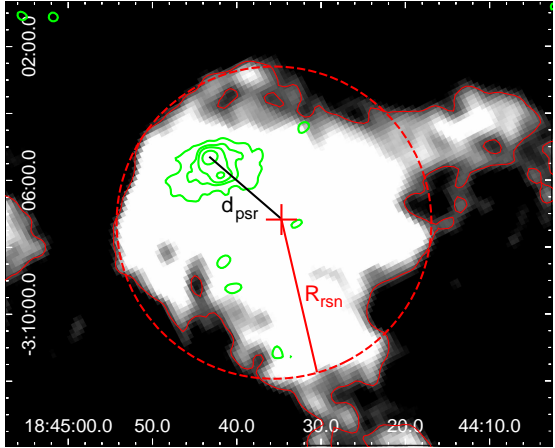}
\caption{Radio continuum emission at 1.4 GHz toward G29.37+0.1 extracted from the MAGPIS (in gray with red contours). 
The brightness scale is saturated to show the boundary of the diffuse emission, approximated by the red dotted circle of radius
$R_{snr}$. X-ray emission from the PWN G29.4+0.1 is shown with green contours, and $d_{psr}$ represents the distance
between the PS1 and the center of the SNR. }
\label{fig_geom}
\end{figure}

We used Eq. \ref{eq1} to estimate the age of the SNR and derive the velocity of the putative pulsar PS1, considering
that they formed together.  
The presence of a pulsar indicates that the SNR formed after the collapse of a massive star that ended in a Type II SN. 
An amount of $E_0 \sim (0.5-4)\times 10^{51}$ erg of mechanical energy is 
expected to be injected into the ISM by a Type II SN; the typical value is $0.9 \times 10^{51}$ erg \citep{kasen09}.
Recently, \citet{muller17} (and references therein) considered a more restrictive upper limit of $2 \times 10^{51}$ erg.
Regarding the ambient density where the SNR expands, in Sect. \ref{dens} we analyzed the HI distribution and obtained an estimation 
of $n_0\sim 8$ cm$^{-3}$ under the assumption that the remnant's blast wave has swept part of the HI detected around it.
In the following analysis we use $n_0=10$~cm$^{-3}$ as an order-of-magnitude approach to a possible value for a dense
medium where the SNR could be expanding.
We note that it is possible that the SN event occurred in an ambient medium that was previously evacuated by the stellar
wind of the progenitor massive star and that the HI shell was modeled by the combined action of both stellar wind and SNR blast wave.
For this reason, we analyze the SNR-PWN system by also exploring the evolution into an ambient medium
with lower densities. 

According to Fig. \ref{fig_geom}, the angular separation between the pulsar PS1 and the center of the SNR is $\sim 2.8^{\prime}$, 
and the radio shell can be approximated by a circle with a radius of $\sim 4.7^{\prime}$.
For a distance of 6.5 kpc, we obtain that the separation between PS1 and the center of the SNR projected in the plane
of the sky is $d_{psr} \sim 5.3$ pc and the radius of the SNR is $R_{rsn} \sim 8.9$ pc.
The pulsar transverse velocity is estimated as $V_{psr}=d_{psr}/t_{snr}$, 
where $d_{psr}$ is the distance between the PSR and the center of the SNR.
This velocity is a good approximation of the spatial velocity of the pulsar if the radial velocity component
of the motion is considerable smaller than the transverse one. Otherwise, it is a lower limit of the 3D velocity of the pulsar.
Considering $n_0=10$ cm$^{-3}$, we obtain $t_{snr} = 20000 - 10000$ yr 
and $V_{psr} = 270 - 530$ km s$^{-1}$ for an SN explosion with energies $E_{51}=0.5-2$, where we have defined $E_{51}=E_0/10^{51}$ erg.
For an ambient medium with $n_0 = 1$ cm$^{-3}$, we obtain $t_{snr} = 6000 - 3000$ yr 
and $V_{psr} = 800 - 1700$ km s$^{-1}$ for an SN explosion with energies $E_{51}=0.5-2$.
For an even lower density medium ($\lesssim 0.1 $ cm$^{-3}$) and still considering a low-energy SN explosion, 
we obtain $V_{psr}>2500$ km s$^{-1}$, which implies an unrealistically fast pulsar when compared with the expected velocity of known pulsars. 
Kick velocities of neutron stars are typically some hundred km s$^{-1}$, with some sources observed to be moving
at $\gtrsim 1000$ km s$^{-1}$ \citep{holland17,verbunt17}. Then, if the SNR is evolving in a low-density medium and independently
of the assume SN explosion energy, an SNR-PWN association is discarded because of the high velocity of the pulsar.
As a reference, in Table \ref{table_sedov} we report the results of the Sedov modeling of the SNR assuming
an SN with a canonical explosion energy $E_{51}=1$ and a distance of 6.5 kpc. 

\begin{table}[h]
\caption{Measured and derived parameters of the SNR-PWN system for a distance of 6.5 kpc. The age of the SNR ($t_{snr}$) is calculated from
Eq. \ref{eq1} considering an ambient density $n_0=1$ and 10 cm$^{-3}$. $P_{snr}$ is the gas pressure inside the SNR (see Sect. \ref{static}).}
\label{table_sedov}
\small
\centering
\setlength{\extrarowheight}{5pt}
\hrule
\begin{tabular}{ll}
\multicolumn{2}{c}{Characteristic sizes/distances, $D=6.5$ kpc} \\
$R_{snr}=8.9$ pc   & Radius of the SNR \\
$R_{pwn}=1.9$ pc   & Radius of the PWN  \\
$d_{psr}=5.3$ pc   & Pulsar - SNR distance \\
\end{tabular}
\hrule
\begin{tabular}{lcc}
\multicolumn{3}{c}{Sedov solution with $E_{0}=10^{51}$ erg}   \\
$n_0$ [cm$^{-3}$]         & 10                  & 1\\
$t_{snr}$ [yr]           & 14000               & 4500 \\
$P_{snr}$ [erg cm$^{-3}$] & $3.6\times 10^{-9}$ & $3.6\times 10^{-9}$      \\
$V_{psr}=d_{psr}/t_{snr}$ [km s$^{-1}$]   & 380                 & 1200 \\
\end{tabular}
\hrule
\end{table}

\subsection{G29.4+0.1, a pulsar wind nebula inside a supernova remnant? }
\label{static}

We now explore the SNR-G29.4+0.1 association using a simple model that describes the evolution of a PWNe inside an SNR.  
During the Sedov stage, the PWN expands subsonically into the remnant ejecta because the interior of the SNR was reheated by 
the reverse shock, increasing the sound speed. \citet{swaluw01} developed an order-of-magnitude analysis of the evolution 
of a PWN deep inside a Sedov SNR based on the equilibrium at the boundary of the PWN 
between the pressure of the wind material ($P_{pwn}$) and the pressure of the gas in the SNR interior ($P_{snr}$).

The Sedov solution yields a 
pressure profile $P(r)$ that is almost uniform for most of the remnant, increasing rapidly only when approaching
the forward shock position at $R_{snr}$ \citep{doku02}. 
We note that G29.4+0.1 is eccentric with respect to the SNR, but if the motion of the pulsar is mostly transversal,  
G29.4+0.1 is located well inside the SNR and completely confined by it, as $R_{pwn}<<R_{snr}$, where $R_{pwn}$ is 
the characteristic size of the PWN. Then, the gas pressure at the position of the PWN can be considered equal 
to the uniform pressure of the inner regions of the SNR ($P_{snr}$).
Following \citet{swaluw01}, we take $P_{snr} = C_1 E_0 / R_{snr}^3\propto t^{-6/5}$, 
with $C_1 \sim 0.074$. For the PWN, the pressure is $P_{pwn} = C_2 L_0 t /R_{pwn}^3 \propto t^{11/15} $, 
with $C_2=(4\pi/3)[(\gamma/(\gamma-1)+6/5]^{(-1/3)} \sim 0.40$ for $\gamma=5/3$ and 
$L_0$ the mechanical luminosity of the wind.
Pressure balance ($P_{snr}=P_{pwn}$) gives the radius of the nebula with respect to the evolutionary state of the remnant:
\begin{equation}
R_{pwn} = \tilde{C} \left(\frac{L_0 t}{E_0} \right)^{1/3} R_{snr}(t),
\label{eq2}
\end{equation}
where $\tilde{C}= C_2/C_1^{1/3} \sim 0.95$. If the rotational energy of the neutron star goes
entirely into the wind, then $L_0 = \dot{E}$, with $\dot{E}$ the spin-down energy of the pulsar. 
The combination of Eqs. \ref{eq1} and \ref{eq2} yields 
$R_{pwn} = \tilde{C} \xi_0^{-5/6}E_0^{-1/2} \dot{E}^{1/3}\rho_0^{1/6}R_{snr}^{11/6}$, i.e. 
$R_{pwn} \propto R_{snr}^{11/6} \propto t^{11/15}$. 
Considering an ISM with 10\% of He, the last equation becomes
\begin{equation}
R_{pwn} = 0.016 E_{51}^{-1/2}\dot{E}_{37}^{1/3}n_3^{1/6}R_{snr}^{11/6},
\label{eq3}
\end{equation}
where $\dot{E}_{37}=\dot{E}/(10^{37}$ erg s$^{-1})$ and the characteristic radii $R_{pwn}$ and $R_{snr}$ are expressed in parsec.  

\begin{figure}[h]
\centering
\includegraphics[width=\linewidth]{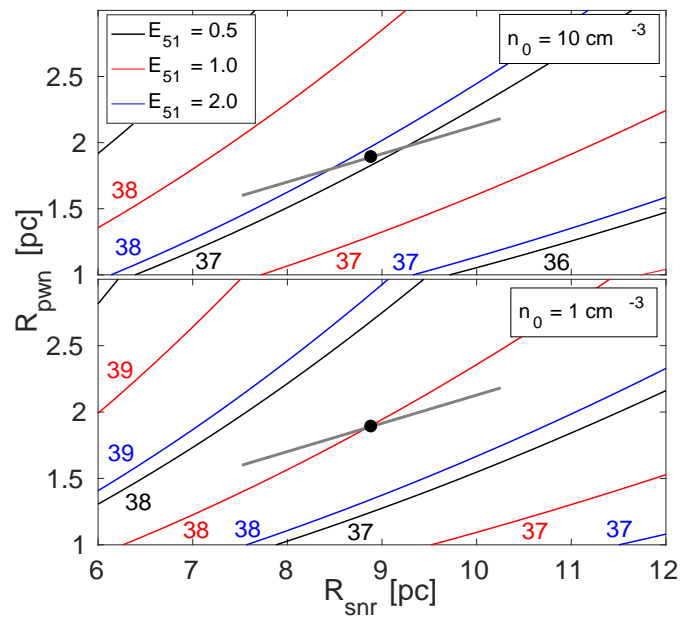}
\caption{Radius of the PWN ($R_{pwn}$) with respect to the evolutionary state of a Sedov SNR, given by
the radius of the forward shock $R_{snr}$ (Eq. \ref{eq3}). We show the results for $n_0=10$ (top) and 1~cm$^{-3}$ (bottom).
The number next to each curve is $log(\dot{E})$, i.e., the spin-down energy of the powering pulsar.
The filled circle is the size of the SNR and the PWN G29.4+0.1 for a distance of 6.5 kpc, while
the dotted line gives $R_{snr}$ and $R_{pwn}$ for the distance range between 5.5 and 7.5 kpc.} 
\label{fig_swaluw}
\end{figure}

Equation \ref{eq3} gives the size of a static PWN inside an SNR in the Sedov stage.
In Fig. \ref{fig_swaluw} we plot this relation for $n_0=10$ and 1~cm$^{-3}$ and different SN explosion energies and pulsar
spin-down energies. The filled circle is the radius of the SNR linked to G29.37+0.1 and the PWN G29.4+0.1 for a 
distance of 6.5 kpc, and the dotted line corresponds 
the same radii evaluated in the $5.5-7.5$ kpc distance range. Because the PWN has an elliptical shape, we approximated its radius
with the mean value of the major and minor semiaxis, which gives $R_{pwn}\sim 1^{\prime}$.    

We analyze the results of Fig. \ref{fig_swaluw} together with the pulsar properties derived in Sect. \ref{lumi}.
We do not intend to use this analysis to determine the SN energy explosion and/or the pulsar energetics, but 
to explore if the SNR-PWN association is possible considering the commonly quoted values for $E_0$ of a core-collapse
SN event and the range of $\dot{E}$ derived from the X-ray properties of G29.4+0.1 and PS1. 
For $n_0 = 10$ cm$^{-3}$, a low-energy SN explosion ($E_{51} = 0.5$) requires a lower energetic 
pulsar ($\dot{E}_{37}\sim 1$), while a high-energy SN explosion ($E_{51}=2$) requires a more energetic
pulsar ($\dot{E}_{37}\sim 9$), slightly above the upper limit of $\dot{E}$ that we defined in Sect. \ref{lumi} for PS1, 
but still acceptable based on the roughness of this analysis. In a lower density medium ($n_0=1$ cm$^{-3}$), 
a low-energy explosion SN ($E_{51}=0.5$) and a pulsar with $\dot{E}_{37} \sim 4 $ are a valid set of parameters to explain the 
confinement of the PWN by the gas inside the SNR. However, a higher energy SN ($E_{51} \gtrsim 1$) requires 
a pulsar spin-down energy $\dot{E}_{37} \gtrsim 10$, which is higher than expected for PS1. 

This analysis shows that it is possible for the pulsar wind to be confined by the internal pressure of the SNR for 
different values of $E_0$ (compatible with that expected in a core-collapse SN event) and $\dot{E}$ (compatible with
the values derived from the X-ray spectral properties of the pulsar and its PWN).
In other words, G29.4+0.1 is a static PWN inside the host SNR, and its distorted morphology in the X-ray band 
could be produced by the pulsar motion toward the left border of the SNR in combination with the asymmetric interaction between 
the pulsar wind and the remnant's reverse shock.

\subsection{Bow-shock PWN inside an SNR}
\label{bow}

We have shown that G29.4+0.1 is likely confined by the gas pressure of the SNR interior. As the powering pulsar is
expected to be moving at several hundred km s$^{-1}$, we investigate if it could be powering a bow-shock PWN in contrast to 
the static scenario.
 
When a fast-moving pulsar is powering a bow shock, the PWN usually acquires a cometary appearance elongated
along the direction of motion, with the pulsar located ahead of the nebula. 
The termination shock turns into a bullet-like structure with strong X-ray and/or radio emission  
surrounded by a fainter tail extending to the back. Neither of these features 
are evident in the X-ray images of G29.4+0.1, which appears elongated perpendicular to the putative direction of motion.
Nevertheless, we explore the bow-shock scenario of G29.4+0.1 by analyzing if the pulsar wind can be balanced
by the ram pressure of the pulsar motion ($P_{ram}$) rather than the pressure of the ambient gas ($P_{amb}$), using the 
system parameters derived in previous sections.  

In the direction of the pulsar motion, the termination shock is referred to as the stand-off distance $R_{w,0}$ and
the ratio between the forward and backward termination shocks gives an estimate of the
Mach number: $\mathcal{M} = v_{r}/c_s \sim R_{w,0}/R_{w,b}$, with $v_r$ the velocity of the pulsar relative to the local medium and 
$c_s$ the local speed of sound.
Low Mach numbers ($\mathcal{M}\gtrsim 1$) are expected in pulsars moving inside SNRs,
while $\mathcal{M} \gg 1$ for pulsars in the ISM \citep{gaensler06}. 
In a few cases, high-resolution X-ray observations in combination with numerical simulations 
revealed the internal structure of bow-shock PWNe and lead to the identification of the termination shock, 
such as the Mouse \citep{gaensler04} and the PWN inside IC 443 \citep{gaensler06b}. 
Lacking the identification of the termination shock, 
we can invoke a scaling relation between $R_{w,0}$ and the forward bow-shock distance $R_f$ 
(i.e., the distance between the pulsar and the apex of the PWN emission): $R_{w,0}\sim 0.5 R_f$ 
(see \citealt{gaensler06} and references therein). 
For a pulsar moving in a medium of density $\rho$, the ram pressure is 
$P_{ram}=\rho v_{r}^2$. Using the sound speed definition $c_s=\sqrt{\gamma P_{amb}/\rho}$, with
$\gamma$ the adiabatic heat ratio, the pressure balance at $R_{w,0}$ between the pulsar wind (assumed to be isotropic)
and the ram pressure is given by
\begin{equation}
\frac{\dot{E}}{4 \pi c R_{w,0}^2} = P_{ram}= \gamma \mathcal{M}^2 P_{amb}.
\label{eq4}
\end{equation}
When we consider that G29.4+0.1 is inside its host SNR at the Sedov stage, 
$P_{amb}=P_{snr}\sim 10^{-9}$ erg cm$^{-3}$. When we take $\dot{E}=10^{37}$ erg s$^{-1}$ and
$\gamma = 5/3$, the bow-shock condition $\mathcal{M}>1$ yields $R_{w,0} \lesssim 0.04$ pc for a distance of 6.5 kpc.
When the scaling relation $R_{w,0}\sim 0.5 R_f$ is adopted, the distance between PS1 and the X-ray bow-shock in the forward direction should be 
$R_f\sim 0.08$~pc. For G29.4+0.1, the angular distance between PS1 and the apex of the X-ray emission is $\sim 0.3^{\prime}$, 
which for a distance of 6.5 kpc yields $R_f \sim 0.6$ pc. This is 
an order of magnitude larger than calculated with Eq. \ref{eq4}.

We conclude that G29.4+0.1 should appear by far more compact than detected in the direction of motion if the pulsar wind were 
confined by the ram pressure of the pulsar moving inside the SNR. This result rules out a bow-shock scenario for G29.4+0.1, although 
we must note that Eq. \ref{eq4} does not take into consideration several factors that can affect the morphology of 
the PWN, such as the viewing angle, a nonisotropic pulsar wind, or inhomogeneities of the surrounding medium \citep{kargal17}.

\subsection{Nature of HESS J1844-030}
\label{hess}

C17 suggested two probable origins for the $\gamma$-rays from HESS J1844-030. 
On the one hand, the point-source morphology of the TeV emission is compatible with 
an extragalactic origin, where the TeV radiation would originate in the lobes of the radio galaxy at the center
of G29.37+0.1. On the other hand, in a Galactic scenario, HESS J1844-030 could be the very high energy counterpart
of the PWN G29.4+0.1 and TeV radiations would originate in a leptonic mechanism. 
C17 did not investigate these scenarios further because by the time they presented their paper, 
HESS J1844-030 had recently been identified as a new TeV source candidate and little information about its spectral properties was available.
We note that in a Galactic context, the TeV source could have an hadronic origin as a consequence of the interaction between 
the SNR and dense material of the ISM. In this case, we would expect a morphological match between the TeV emission 
and the distribution of the neutral or molecular gas. Several molecular clouds (identified by C17) and an HI arm (see Sect. \ref{dens}) 
are found in the direction of G29.37+0.1, but in none of them is the most intense emission detected coincident with the centroid of the TeV 
radiation from HESS J1844-030. 

In the context of our study of the PWN G29.4+0.1 and its pulsar, we examined one of the scenarii described above: is
HESS J1844-030 the very high energy counterpart of the X-ray PWN? 
TeV sources are classified as PWN based on their spatial and/or morphological coincidence with a PWN detected in other spectral band or based
on their energy-dependent TeV morphology. Using data from the latest HGPS, \citet{abdalla18pwn} reported 14 confirmed and 10 candidate TeV
PWNe detected by HESS. Most of the TeV PWNe are detected as extended sources in the very high energy domain and
present spatial variations of the spectrum, which is expected to become softer with increasing distance from the powering pulsar. 
Within the population of confirmed HESS PWNe, only 2 sources do not show an extended morphology in the TeV band. 
Interestingly, in one of them (HESS J1833-105), the lack of extension in the TeV domains is interpreted as evidence 
that the $\gamma$-ray emission is produced by the pulsar wind rather than by the front shock of the host SNR interacting with the 
ambient medium. This means that the point-source nature of HESS J1844-030 does not rule out a PWN origin, and we investigate if its
spectral properties in the TeV domain are similar to those found toward the population of Galactic PWNe.
In Table \ref{table_disc} we report the spectral properties of the PWN G29.4+0.1 and HESS J1844-030 that we used to 
analyze a possible connection between them. 

\begin{table}[h]
\caption{Spectral properties of PWN G29.4+0.1 ({\it pwn}), the pulsar PS1 ({\it psr}), and HESS J1844-030 ({\it TeV}) for a 
distance of 6.5 kpc. The parameters are obtained from the spectral analysis
of Sects. \ref{spec_todo} and \ref{spec_anillos}, except for $\Gamma_{TeV}$ and the flux in the TeV band
(used to estimate $L_{TeV}$), which were taken from \citet{abdalla18}.}
\label{table_disc}
\small
\centering
\setlength{\extrarowheight}{5pt}
\hrule
\begin{tabular}{ll}
$\Gamma_{pwn}=1.87$                        & PWN photon index in the keV band \\
$L_{pwn}=1.16 \times 10^{34}$ erg s$^{-1}$ & PWN luminosity in $0.5 - 8.0$ keV \\
$\Gamma_{psr}=1.35$                        & Pulsar photon index in the keV band \\
$L_{psr}=6.13 \times 10^{32}$ erg s$^{-1}$ & Pulsar luminosity $0.5 - 8.0$ keV \\
$\Gamma_{TeV}=2.48$                        & TeV spectral index  \\
$L_{TeV}=4.61 \times 10^{33}$ erg s$^{-1}$ & Luminosity in $1 - 10$ TeV \\
$d_{psr-TeV}= 1$ pc                        & Pulsar - TeV centroid distance\\ 
\end{tabular}
\hrule
\end{table}

According to the HGPS catalog, HESS J1843-030 is detected as a point source 
in the TeV domain, and the best-fit position for a single-Gaussian model yields a centroid $l = 29.41 \pm 0.01; b=0.09 \pm 0.01$. 
A power-law fitting gives a best-fit spectral index 
$\Gamma_{TeV}\sim 2.48$ and an energy flux of $\sim 9.12 \times 10^{-13}$ erg cm$^{-2}$ s$^{-1}$ in the $1-10$ TeV energy band. 
This translates into a luminosity $L_{TeV}\sim 4.6 \times 10^{33}$ erg s$^{-1}$ in the $1-10$ TeV energy band for a distance of 
6.5 kpc. 
Taking the spin-down power of the pulsar PS1 powering G29.4+0.1 $\dot{E}\sim 10^{37}$ erg s$^{-1}$ (see Sect. \ref{lumi}),
we obtain a TeV efficiency $\eta_{TeV} = L_{TeV} / \dot{E} \sim 5 \times 10^{-4}$.
The projected separation between PS1 and the TeV centroid is $\sim 0.6^{\prime}$, which translates into a pulsar offset of 
$d_{psr-TeV}\sim 1$ pc for the distance considered above.
\citet{abdalla18pwn} reported a large dispersion in the $L_{TeV} - \dot{E}$ relation for the population of 
PWNe detected by HESS (both confirmed and candidate) and an apparent correlation between $\eta_{TeV}$ and $d_{psr-TeV}$.
The locations of the HESS J1844-030/PS1 system in the $L_{TeV} - \dot{E}$ and $\eta_{TeV} - d_{psr-TeV}$ diagrams
of \citet{abdalla18pwn} (see their Figs. 7 and 9) are similar to those of other HESS PWN/pulsar.
On the other hand, \citet{kargal13} studied the population of Galactic PWNe in the keV and TeV domains to search for 
possible correlations between luminosities and/or spectral indexes. They find large scatter in their
$L_X - \dot{E}$ (as described in Sect. \ref{lumi}), $L_X - L_{TeV}$ and $L_{TeV} - \Gamma_{TeV}$ diagrams, with
$L_X$ the PWN luminosity in the $0.5-8.0$ keV band. 
The behavior of $L_X$, $L_{TeV}$ and $\Gamma_{TeV}$ for the G29.4+0.1/HESS J1844-030 system is similar 
to that observed in other keV/TeV Galactic PWNe.

The overall spectral properties of the PWN G29.4+0.1, the pulsar, and HESS J1844-030 in the keV and TeV bands are similar
to other PWNe. However, based on the large dispersion of the parameters observed toward similar Galactic systems, 
we conclude that a PWN origin for HESS J1844-030 cannot be confirmed or discarded.
The possible connection between G29.4+01 and the HESS source can be investigated considering the underlying emission mechanisms
responsible for the X- and $\gamma$-ray emissions of PWNe.
If HESS J1844-030 is the very high energy counterpart of G29.4+0.1, the TeV emission is expected
to arise from IC scattering of the ambient low-energy photons with the same population of electron producing synchrotron
radiation in the keV band. Then, we expect the photon index in the keV band $\Gamma_X$ to be similar to
the spectral index in the TeV band $\Gamma_{TeV}$ \citep{pavlov08}. 
The TeV spectrum of HESS J1844-030 is considerably softer ($\Gamma_{TeV}=2.5 \pm 0.1$, \citealt{abdalla18}) than the 
X-ray spectrum of G29.4+0.1 ($\Gamma_X=1.87 \pm 0.30$). This behavior has been observed in many PWNe with TeV and X-ray
and \citet{kargal13} discussed probable causes. Following their paper, 
we calculated the energy of the electrons producing synchrotron photons in the $1.5-8.0$ keV band as 
$E_e \sim 160E_{keV}^{1/2}B_{\mu\rm{G}}^{-1/2} \rm{TeV}$. For $B=5.6$ $\mu$G, we obtain $E_e \sim 80-190~\rm{TeV}$.  
When this population of relativistic electrons interact with background photons of energy $\epsilon$ via IC 
scattering, they produce $\gamma$-ray photons in the TeV band.
We estimate the energy of the TeV photons as $E_{IC} \sim 4 (E_e/\rm{TeV})^2 \epsilon_{eV}~\rm{TeV}$,
where $\epsilon_{eV} = \epsilon / (1~\rm{eV})$.
If the CMB is the main source of background photons, then $\epsilon \sim 3 \times 10^{-4}$ eV (for $T_{CMB} \sim 3$ K). 
For $E_e \sim 80 - 190 $ TeV, the energy of the $\gamma$-ray photons produced by the IC scattering is $E_{IC} \sim 8 - 40$ TeV.
$\gamma$-ray emission from HESS J1844-030 is detected up to 50 TeV \citep{abdalla18}, 
showing that with the nebular magnetic field obtained in previous sections, it is possible for the 
same population of electrons to power keV emission from G29.4+0.1 and TeV emission from HESS J1844-030.  

Another possible scenario for HESS J1844-030 has been proposed recently.
C17 first noted that the X-ray point source PS2 
presents strong flux variability, low absorption in the keV band (pointing to a Galactic source), and a 
hard nonthermal X-ray spectrum.
Based on these characteristics, together with the excellent positional match between PS2
and the TeV centroid and the point-source nature of the TeV source, \citet{mccall18} 
suggested that HESS J1844-030 could be a new $\gamma$-ray binary, 
whose counterpart in the keV band is PS2. 
Currently, six of these sources are detected at GeV or TeV energies. They 
are characterized by periodic (and some of them also sporadic) variability that is modulated by the orbital period of the system,
and they display hard X-ray ($\Gamma_X < 2$) and softer TeV spectra \citep{dubus17,aliu14}.
Based on the hard X-ray spectrum of PS2 ($\Gamma_X < 1$, C17) and the soft spectrum
of HESS J1844-030 ($\Gamma_{TeV} \sim 2.5$, \citealt{abdalla18}), this scenario cannot be discarded, although the 
detection of variability in the $\gamma$-ray domain is necessary to confirm it. 

\section{Conclusions}
\label{concl}

We now summarize and discuss the results of our analysis. We also point out the assumptions we have 
taken into consideration.  

The spectral characteristics of the nebula and the point source PS1 in the X-ray band (photon indexes, 
luminosities, and the softening of the spectrum of the diffuse emission) are evidence that G29.4+0.1 is likely a PWN powered by 
the pulsar PS1.
We suggest that part of the radio emission from the`S-shaped feature (associated with a 
radio galaxy) could arise from the relic PWN left behind by the pulsar motion.
Polarimetric observations could help to detect such  a nebula based on the high degree of polarization of the radio continuum emission
from PWNe.
We used empirical relations reported in the literature to obtain a crude estimate of the pulsar's spin-down power, obtaining
$\dot{E}\sim 10^{37}$ erg s$^{-1}$. 
From the analysis of the HI absorption spectra we set a distance range of $5.5 - 7.5$ kpc 
for the SNR linked to the radio source G29.37+0.1. The distribution of the neutral gas showed that the ISM is nonuniform,
presenting a higher-density medium to the left of the SNR. We identified an arm of HI at 6.5 kpc that delineates part of the radio emission, and 
we speculated that this material could have been swept by the blast wave of the SNR.
We analyzed the connection between the PWN G29.4+0.1 and the SNR and found that the characteristic sizes of both PWN and SNR are
compatible with the pulsar wind being confined by the gas pressure inside the remnant, under the assumption that
it is in the Sedov stage of evolution. The distorted morphology of the nebula in the X-ray band, with the pulsar ahead of the
emission in the direction of the expected motion from the center of the SNR, could be caused by the asymmetric interaction
of the PWN with the remnant's reverse shock in combination with the motion of the pulsar.
A bow-shock scenario for the PWN is discarded as the X-ray nebula should appear far more compact
than is observed in the direction of motion of the putative pulsar. 
Finally, using the magnetic field derived from the X-ray luminosity, we showed that the population 
of relativistic electrons producing synchrotron radiation in the keV band from G29.4+0.1 can also produce
TeV photons through the IC mechanism.
This order-of-magnitude analysis suggests an association between G29.4+0.1 and HESS J1844-030, although 
a broadband spectral analysis (from radio to TeV) is required to confirm this scenario among other possible ones.

One of the assumptions of our analysis is that the current evolutionary stage of the SNR can be modeled by the Sedov solution
in a homogeneous medium of density $n_0$.  
The condition of homogeneity of the surrounding medium where the blast wave propagates can be questioned based on the 
distribution of the neutral atomic gas presented in Sect. \ref{dens}. The SNR appears to be located in the interface between 
higher (to the left) and lower (to the right) density media. 
We speculate that for this reason, the left border of the radio continuum emission appears
sharper and flatter than the right border, where the radio emission fades irregularly. Nevertheless, the overall shape
of the SNR projected in the plane of the sky presents a quite circular shape, indicating that the interaction with the ISM
has not distorted its morphology considerably and the condition of homogeneity of the ambient medium could still be considered 
as valid. 

Another issue we examined is the value of the ISM density. 
The Sedov modeling of the evolution of a SNR depends on the ambient density $n_0$: lower values of $n_0$ yield 
a younger SNR. For G29.37+0.1, the HI analysis showed a structure of neutral gas with 
an arm-like shape that delineates the bottom part of the SNR. Assuming that
the blast wave has swept this material, we obtained a rough estimate of $n_0 \sim 10$~cm$^{-3}$.
We point out that the left border of the SNR (the one that appears flatter in the radio continuum observations) is seen superimposed 
to the large HI structure, so that part of this neutral material could have been shocked by the blast wave. We did not include it in the 
calculation of $n_0$ because it does not present any conspicuous structure (such as a shell or clump) that could help to 
separate the shocked from the unshocked HI. On the other hand, if the SN exploded in a medium that was previously evacuated by the stellar
winds of the progenitor star, the SNR would be evolving in a low-density medium. 
For this reason, we modeled the evolution of the SNR in the Sedov stage considering
two possible densities values, $n_0 =10$ and 1 cm$^{-3}$. 
As we explored the possible connection between the SNR and the PWN, an even lower density medium was discarded because it yielded
a quite young SNR, which requires an unrealistically fast pulsar to reconcile an SNR-PWN common origin.  

Another assumption is related to the 3D motion of the pulsar inside the SNR. We do not have any clue on its radial velocity, 
therefore we considered a pure transverse motion. As a consequence, the distances and sizes associated with the PWN-SNR
system are taken equal to their projected values on the plane of the sky.
In this scenario, the distance between the pulsar and the center of the 
SNR is $\sim 0.6 R_{snr}$, where $R_{snr}$ is the radius of the SNR. Thus, the pulsar is still located well inside the SNR
and the gas pressure of the remnant at the position of G29.4+0.1 is $\propto E_0/R_{snr}^3$ , as obtained from the Sedov solution 
adopted in our analysis. If the pulsar had a considerable radial motion, it could be close to 
the remnant's shock front where the pressure ceases to be uniform and rapidly increases. 
 
We recall that our analysis is not intended to determine the unknown properties of the system
(such as the SN explosion energy or the pulsar spin-down power), but to explore if a connection between G29.4+0.1 and
the suspected SNR is possible considering a set of parameter values that are commonly found in PWNe, pulsars, SNRs and in the Galactic ISM.
Some of these parameters are usually constrained by observations. For instance, pulsed emission from neutron stars may be suitable
to derive its spin-down power, characteristic age, and distance (from its dispersion measure).
This would allow a comparison with the age and distance of the SNR derived from the Sedov model and the HI analysis, respectively, 
to support an association between them. So far, no pulsations from PS1 are detected in any spectral {\it band}, 
and this can be a consequence of 
the limited sensitivity of the observations or the fact that the beam of the pulsar does not point toward the Earth.

\begin{acknowledgements}
A. Petriella wishes to acknowledge the comments and suggestions of the anonymous referee that improved the paper considerably.
A. Petriella is a member of the Carrera del Investigador Cient\'ifico
of CONICET, Argentina. This work was partially supported by Argentina grants awarded by UBA (UBA-CyT), CONICET and ANPCYT.
This research has made use of data obtained from the Chandra Data Archive and software provided by the 
Chandra X-ray Center (CXC) in the application packages CIAO, ChIPS, and Sherpa.
This research is partially based on observations obtained with XMM-Newton, an ESA science mission with instruments
and contributions directly funded by ESA Member States and NASA.
\end{acknowledgements}

\bibliographystyle{aa} 
\bibliography{biblio} 

\begin{thebibliography}{63}
\expandafter\ifx\csname natexlab\endcsname\relax\def\natexlab#1{#1}\fi

\bibitem[{{Aliu} {et~al.}(2014){Aliu}, {Archambault}, {Aune}, {Behera},
  {Beilicke}, {Benbow}, {Berger}, {Bird}, {Bouvier}, {Buckley}, \&
  et~al.}]{aliu14}
{Aliu}, E., {Archambault}, S., {Aune}, T., {et~al.} 2014, \apj, 780, 168

\bibitem[{{Blondin} {et~al.}(2001){Blondin}, {Chevalier}, \&
  {Frierson}}]{blondin01}
{Blondin}, J.~M., {Chevalier}, R.~A., \& {Frierson}, D.~M. 2001, \apj, 563, 806

\bibitem[{{Bocchino} \& {Bykov}(2001)}]{bocchino01b}
{Bocchino}, F. \& {Bykov}, A.~M. 2001, \aap, 376, 248

\bibitem[{{Bocchino} {et~al.}(2001){Bocchino}, {Warwick}, {Marty}, {Lumb},
  {Becker}, \& {Pigot}}]{bocchino01a}
{Bocchino}, F., {Warwick}, R.~S., {Marty}, P., {et~al.} 2001, \aap, 369, 1078

\bibitem[{{Bosch-Ramon} {et~al.}(2010){Bosch-Ramon}, {Romero}, {Araudo}, \&
  {Paredes}}]{bosch-ramon10}
{Bosch-Ramon}, V., {Romero}, G.~E., {Araudo}, A.~T., \& {Paredes}, J.~M. 2010,
  \aap, 511, A8

\bibitem[{{Camilo} {et~al.}(2006){Camilo}, {Ransom}, {Gaensler}, {Slane},
  {Lorimer}, {Reynolds}, {Manchester}, \& {Murray}}]{camilo06}
{Camilo}, F., {Ransom}, S.~M., {Gaensler}, B.~M., {et~al.} 2006, \apj, 637, 456

\bibitem[{{Caraveo} {et~al.}(2003){Caraveo}, {Bignami}, {De Luca},
  {Mereghetti}, {Pellizzoni}, {Mignani}, {Tur}, \& {Becker}}]{caraveo03}
{Caraveo}, P.~A., {Bignami}, G.~F., {De Luca}, A., {et~al.} 2003, Science, 301,
  1345

\bibitem[{{Castelletti} {et~al.}(2007){Castelletti}, {Dubner}, {Brogan}, \&
  {Kassim}}]{caste07}
{Castelletti}, G., {Dubner}, G., {Brogan}, C., \& {Kassim}, N.~E. 2007, \aap,
  471, 537

\bibitem[{{Castelletti} {et~al.}(2011){Castelletti}, {Dubner}, {Clarke}, \&
  {Kassim}}]{caste11}
{Castelletti}, G., {Dubner}, G., {Clarke}, T., \& {Kassim}, N.~E. 2011, \aap,
  534, A21

\bibitem[{{Castelletti} {et~al.}(2017){Castelletti}, {Supan}, {Petriella},
  {Giacani}, \& {Joshi}}]{caste17}
{Castelletti}, G., {Supan}, L., {Petriella}, A., {Giacani}, E., \& {Joshi},
  B.~C. 2017, \aap, 602, A31

\bibitem[{{Ciotti} \& {D'Ercole}(1989)}]{ciotti89}
{Ciotti}, L. \& {D'Ercole}, A. 1989, \aap, 215, 347

\bibitem[{{de Naurois}(2015)}]{naurois15}
{de Naurois}, M. 2015, in International Cosmic Ray Conference, Vol.~34, 34th
  International Cosmic Ray Conference (ICRC2015), 21

\bibitem[{{Dokuchaev}(2002)}]{doku02}
{Dokuchaev}, V.~I. 2002, \aap, 395, 1023

\bibitem[{{Dubner} {et~al.}(2002){Dubner}, {Gaensler}, {Giacani}, {Goss}, \&
  {Green}}]{dubner02}
{Dubner}, G.~M., {Gaensler}, B.~M., {Giacani}, E.~B., {Goss}, W.~M., \&
  {Green}, A.~J. 2002, \aj, 123, 337

\bibitem[{{Dubus} {et~al.}(2017){Dubus}, {Guillard}, {Petrucci}, \&
  {Martin}}]{dubus17}
{Dubus}, G., {Guillard}, N., {Petrucci}, P.-O., \& {Martin}, P. 2017, \aap,
  608, A59

\bibitem[{{Ferrand} \& {Safi-Harb}(2012)}]{ferrand12}
{Ferrand}, G. \& {Safi-Harb}, S. 2012, Advances in Space Research, 49, 1313

\bibitem[{{Fich} {et~al.}(1989){Fich}, {Blitz}, \& {Stark}}]{fich89}
{Fich}, M., {Blitz}, L., \& {Stark}, A.~A. 1989, \apj, 342, 272

\bibitem[{{Franco} {et~al.}(1991){Franco}, {Tenorio-Tagle}, {Bodenheimer}, \&
  {Rozyczka}}]{franco91}
{Franco}, J., {Tenorio-Tagle}, G., {Bodenheimer}, P., \& {Rozyczka}, M. 1991,
  \pasp, 103, 803

\bibitem[{{Gaensler} {et~al.}(1999){Gaensler}, {Brazier}, {Manchester},
  {Johnston}, \& {Green}}]{gaensler99}
{Gaensler}, B.~M., {Brazier}, K.~T.~S., {Manchester}, R.~N., {Johnston}, S., \&
  {Green}, A.~J. 1999, \mnras, 305, 724

\bibitem[{{Gaensler} {et~al.}(2006){Gaensler}, {Chatterjee}, {Slane}, {van der
  Swaluw}, {Camilo}, \& {Hughes}}]{gaensler06b}
{Gaensler}, B.~M., {Chatterjee}, S., {Slane}, P.~O., {et~al.} 2006, \apj, 648,
  1037

\bibitem[{{Gaensler} \& {Slane}(2006)}]{gaensler06}
{Gaensler}, B.~M. \& {Slane}, P.~O. 2006, \araa, 44, 17

\bibitem[{{Gaensler} {et~al.}(2000){Gaensler}, {Stappers}, {Frail}, {Moffett},
  {Johnston}, \& {Chatterjee}}]{gaensler00}
{Gaensler}, B.~M., {Stappers}, B.~W., {Frail}, D.~A., {et~al.} 2000, \mnras,
  318, 58

\bibitem[{{Gaensler} {et~al.}(2004){Gaensler}, {van der Swaluw}, {Camilo},
  {Kaspi}, {Baganoff}, {Yusef-Zadeh}, \& {Manchester}}]{gaensler04}
{Gaensler}, B.~M., {van der Swaluw}, E., {Camilo}, F., {et~al.} 2004, \apj,
  616, 383

\bibitem[{{Gotthelf}(2003)}]{gotthelf03}
{Gotthelf}, E.~V. 2003, \apj, 591, 361

\bibitem[{{Green}(2017)}]{green17}
{Green}, D.~A. 2017, A Catalogue of Galactic Supernova Remnants (2017 June
  version), Cavendish Laboratory, Cambridge, United Kingdom (available at
  http://www.mrao.cam.ac.uk/surveys/snrs/)

\bibitem[{{H.~E.~S.~S.~Collaboration}
  {et~al.}(2018{\natexlab{a}}){H.~E.~S.~S.~Collaboration}, {Abdalla},
  {Abramowski}, {Aharonian}, {Ait Benkhali}, {Akhperjanian}, {Andersson},
  {Ang{\"u}ner}, {Arrieta}, {Aubert}, \& et~al.}]{abdalla18pwn}
{H.~E.~S.~S.~Collaboration}, {Abdalla}, H., {Abramowski}, A., {et~al.}
  2018{\natexlab{a}}, \aap, 612, A2

\bibitem[{{H.~E.~S.~S.~Collaboration}
  {et~al.}(2018{\natexlab{b}}){H.~E.~S.~S.~Collaboration}, {Abdalla},
  {Abramowski}, {Aharonian}, {Benkhali}, {Ang{\"u}ner}, {Arakawa}, {Arrieta},
  {Aubert}, {Backes}, \& et~al.}]{abdalla18}
{H.~E.~S.~S.~Collaboration}, {Abdalla}, H., {Abramowski}, A., {et~al.}
  2018{\natexlab{b}}, \aap, 612, A1

\bibitem[{{Helfand} {et~al.}(2006){Helfand}, {Becker}, {White}, {Fallon}, \&
  {Tuttle}}]{helfand06}
{Helfand}, D.~J., {Becker}, R.~H., {White}, R.~L., {Fallon}, A., \& {Tuttle},
  S. 2006, \aj, 131, 2525

\bibitem[{{Helfand} {et~al.}(1989){Helfand}, {Velusamy}, {Becker}, \&
  {Lockman}}]{helfand89}
{Helfand}, D.~J., {Velusamy}, T., {Becker}, R.~H., \& {Lockman}, F.~J. 1989,
  \apj, 341, 151

\bibitem[{{Holland-Ashford} {et~al.}(2017){Holland-Ashford}, {Lopez},
  {Auchettl}, {Temim}, \& {Ramirez-Ruiz}}]{holland17}
{Holland-Ashford}, T., {Lopez}, L.~A., {Auchettl}, K., {Temim}, T., \&
  {Ramirez-Ruiz}, E. 2017, \apj, 844, 84

\bibitem[{{Holler} {et~al.}(2012){Holler}, {Sch{\"o}ck}, {Eger},
  {Kie{\ss}ling}, {Valerius}, \& {Stegmann}}]{holler12}
{Holler}, M., {Sch{\"o}ck}, F.~M., {Eger}, P., {et~al.} 2012, \aap, 539, A24

\bibitem[{{Jackson} {et~al.}(2006){Jackson}, {Rathborne}, {Shah}, {Simon},
  {Bania}, {Clemens}, {Chambers}, {Johnson}, {Dormody}, {Lavoie}, \&
  {Heyer}}]{jackson06}
{Jackson}, J.~M., {Rathborne}, J.~M., {Shah}, R.~Y., {et~al.} 2006, \apjs, 163,
  145

\bibitem[{{Johanson} \& {Kerton}(2009)}]{johanson09}
{Johanson}, A.~K. \& {Kerton}, C.~R. 2009, \aj, 138, 1615

\bibitem[{{Kargaltsev} \& {Pavlov}(2008)}]{kargal08}
{Kargaltsev}, O. \& {Pavlov}, G.~G. 2008, in American Institute of Physics
  Conference Series, Vol. 983, 40 Years of Pulsars: Millisecond Pulsars,
  Magnetars and More, ed. C.~{Bassa}, Z.~{Wang}, A.~{Cumming}, \& V.~M.
  {Kaspi}, 171--185

\bibitem[{{Kargaltsev} {et~al.}(2017){Kargaltsev}, {Pavlov}, {Klingler}, \&
  {Rangelov}}]{kargal17}
{Kargaltsev}, O., {Pavlov}, G.~G., {Klingler}, N., \& {Rangelov}, B. 2017,
  Journal of Plasma Physics, 83, 635830501

\bibitem[{{Kargaltsev} {et~al.}(2013){Kargaltsev}, {Rangelov}, \&
  {Pavlov}}]{kargal13}
{Kargaltsev}, O., {Rangelov}, B., \& {Pavlov}, G.~G. 2013, ArXiv e-prints
  [\eprint[arXiv]{1305.2552}]

\bibitem[{{Kasen} \& {Woosley}(2009)}]{kasen09}
{Kasen}, D. \& {Woosley}, S.~E. 2009, \apj, 703, 2205

\bibitem[{{Landecker} {et~al.}(1999){Landecker}, {Routledge}, {Reynolds},
  {Smegal}, {Borkowski}, \& {Seward}}]{landecker99}
{Landecker}, T.~L., {Routledge}, D., {Reynolds}, S.~P., {et~al.} 1999, \apj,
  527, 866

\bibitem[{{Ma} {et~al.}(2016){Ma}, {Ng}, {Bucciantini}, {Slane}, {Gaensler}, \&
  {Temim}}]{ma16}
{Ma}, Y.~K., {Ng}, C.-Y., {Bucciantini}, N., {et~al.} 2016, \apj, 820, 100

\bibitem[{{McCall} \& {Errando}(2018)}]{mccall18}
{McCall}, H. \& {Errando}, M. 2018, in American Astronomical Society Meeting
  Abstracts, Vol. 231, American Astronomical Society Meeting Abstracts, 434.02

\bibitem[{{M{\"u}ller}(2017)}]{muller17}
{M{\"u}ller}, B. 2017, in IAU Symposium, Vol. 329, The Lives and Death-Throes
  of Massive Stars, ed. J.~J. {Eldridge}, J.~C. {Bray}, L.~A.~S. {McClelland},
  \& L.~{Xiao}, 17--24

\bibitem[{{Ng} {et~al.}(2008){Ng}, {Slane}, {Gaensler}, \& {Hughes}}]{ng08}
{Ng}, C.-Y., {Slane}, P.~O., {Gaensler}, B.~M., \& {Hughes}, J.~P. 2008, \apj,
  686, 508

\bibitem[{{Park} {et~al.}(2013){Park}, {Koo}, {Gibson}, {Kang}, {Lane},
  {Douglas}, {Peek}, {Korpela}, {Heiles}, \& {Newton}}]{park13}
{Park}, G., {Koo}, B.-C., {Gibson}, S.~J., {et~al.} 2013, \apj, 777, 14

\bibitem[{{Pavlov} {et~al.}(2008){Pavlov}, {Kargaltsev}, \&
  {Brisken}}]{pavlov08}
{Pavlov}, G.~G., {Kargaltsev}, O., \& {Brisken}, W.~F. 2008, \apj, 675, 683

\bibitem[{{Petre} {et~al.}(2002){Petre}, {Kuntz}, \& {Shelton}}]{petre02}
{Petre}, R., {Kuntz}, K.~D., \& {Shelton}, R.~L. 2002, \apj, 579, 404

\bibitem[{{Porquet} {et~al.}(2003){Porquet}, {Decourchelle}, \&
  {Warwick}}]{porquet03}
{Porquet}, D., {Decourchelle}, A., \& {Warwick}, R.~S. 2003, \aap, 401, 197

\bibitem[{{Possenti} {et~al.}(2002){Possenti}, {Cerutti}, {Colpi}, \&
  {Mereghetti}}]{possenti02}
{Possenti}, A., {Cerutti}, R., {Colpi}, M., \& {Mereghetti}, S. 2002, \aap,
  387, 993

\bibitem[{{Ranasinghe} \& {Leahy}(2017)}]{rana17}
{Ranasinghe}, S. \& {Leahy}, D.~A. 2017, \apj, 843, 119

\bibitem[{{Ranasinghe} \& {Leahy}(2018)}]{rana18}
{Ranasinghe}, S. \& {Leahy}, D.~A. 2018, \aj, 155, 204

\bibitem[{{Roman-Duval} {et~al.}(2009){Roman-Duval}, {Jackson}, {Heyer},
  {Johnson}, {Rathborne}, {Shah}, \& {Simon}}]{duval09}
{Roman-Duval}, J., {Jackson}, J.~M., {Heyer}, M., {et~al.} 2009, \apj, 699,
  1153

\bibitem[{{Safi-Harb} {et~al.}(2001){Safi-Harb}, {Harrus}, {Petre}, {Pavlov},
  {Koptsevich}, \& {Sanwal}}]{safi01}
{Safi-Harb}, S., {Harrus}, I.~M., {Petre}, R., {et~al.} 2001, \apj, 561, 308

\bibitem[{{Seward} \& {Wang}(1988)}]{seward88}
{Seward}, F.~D. \& {Wang}, Z.-R. 1988, \apj, 332, 199

\bibitem[{{Stil} {et~al.}(2006){Stil}, {Taylor}, {Dickey}, {Kavars}, {Martin},
  {Rothwell}, {Boothroyd}, {Lockman}, \& {McClure-Griffiths}}]{stil06}
{Stil}, J.~M., {Taylor}, A.~R., {Dickey}, J.~M., {et~al.} 2006, \aj, 132, 1158

\bibitem[{{Sup{\'a}n} {et~al.}(2018){Sup{\'a}n}, {Castelletti}, {Peters}, \&
  {Kassim}}]{supan18}
{Sup{\'a}n}, L., {Castelletti}, G., {Peters}, W.~M., \& {Kassim}, N.~E. 2018,
  ArXiv e-prints [\eprint[arXiv]{1806.07452}]

\bibitem[{{Swartz} {et~al.}(2015){Swartz}, {Pavlov}, {Clarke}, {Castelletti},
  {Zavlin}, {Bucciantini}, {Karovska}, {van der Horst}, {Yukita}, \&
  {Weisskopf}}]{swartz15}
{Swartz}, D.~A., {Pavlov}, G.~G., {Clarke}, T., {et~al.} 2015, \apj, 808, 84

\bibitem[{{Temim} {et~al.}(2015){Temim}, {Slane}, {Kolb}, {Blondin}, {Hughes},
  \& {Bucciantini}}]{temim15}
{Temim}, T., {Slane}, P., {Kolb}, C., {et~al.} 2015, \apj, 808, 100

\bibitem[{{van der Swaluw} {et~al.}(2001){van der Swaluw}, {Achterberg},
  {Gallant}, \& {T{\'o}th}}]{swaluw01}
{van der Swaluw}, E., {Achterberg}, A., {Gallant}, Y.~A., \& {T{\'o}th}, G.
  2001, \aap, 380, 309

\bibitem[{{van der Swaluw} {et~al.}(2004){van der Swaluw}, {Downes}, \&
  {Keegan}}]{swaluw04}
{van der Swaluw}, E., {Downes}, T.~P., \& {Keegan}, R. 2004, \aap, 420, 937

\bibitem[{{Vel{\'a}zquez} {et~al.}(2002){Vel{\'a}zquez}, {Dubner}, {Goss}, \&
  {Green}}]{velazquez02}
{Vel{\'a}zquez}, P.~F., {Dubner}, G.~M., {Goss}, W.~M., \& {Green}, A.~J. 2002,
  \aj, 124, 2145

\bibitem[{{Verbunt} {et~al.}(2017){Verbunt}, {Igoshev}, \& {Cator}}]{verbunt17}
{Verbunt}, F., {Igoshev}, A., \& {Cator}, E. 2017, \aap, 608, A57

\bibitem[{{Vink}(2012)}]{vink12}
{Vink}, J. 2012, \aapr, 20, 49

\bibitem[{{Xiao} {et~al.}(2009){Xiao}, {Reich}, {F{\"u}rst}, \& {Han}}]{xiao09}
{Xiao}, L., {Reich}, W., {F{\"u}rst}, E., \& {Han}, J.~L. 2009, \aap, 503, 827

\bibitem[{{Xiao} \& {Zhu}(2012)}]{xiao12}
{Xiao}, L. \& {Zhu}, M. 2012, \aap, 545, A86

\end{thebibliography}

\end{document}